\documentclass[aps,prd,twocolumn,nofootinbib,groupedaddress,showpacs,amsmath,amssymb,floatfix]{revtex4-2}

\usepackage{physics}
\usepackage{graphicx}
\usepackage[colorlinks=true,allcolors=blue]{hyperref}%

\usepackage{tikz}
\tikzset{every picture/.style={line width=0.25mm}}
\definecolor{pred}{RGB}{238,28,37}
\definecolor{pblue}{RGB}{48,49,146}
\definecolor{pgreen}{RGB}{00,163,80}
\usetikzlibrary{decorations.pathmorphing}

\newcommand{\kk}{\kappa}

\newcommand{\Dvec}[1]{\underline{#1}}

\newcommand{\Laplace}{{\triangle}}

\begin{document}

\newcommand{\UB}{Fakult\"at f\"ur Physik, Universit\"at Bielefeld, D-33615 Bielefeld, Germany}

\title{Regularization of stochastic fluid dynamics and non-perturbative renormalization of the shear viscosity of incompressible fluids}

\author{Mattis Harhoff}
\author{Cedric Riedel}
\author{Johannes Roth}
\author{S\"oren Schlichting}
\affiliation{\UB}

\date{\today}

\begin{abstract} 
We propose a regularization of stochastic fluid dynamics that preserves all relevant symmetries. By considering an incompressible fluid, we demonstrate explicitly that the regularized theory is UV finite, and therefore admits a continuum limit. Strikingly, the regularized theory of stochastic fluid dynamics is not only amenable to direct numerical simulations of the lattice discretized theory, but also to functional methods in the continuum. We exploit this to perform renormalization group improved perturbative calculations of the renormalization of the shear viscosity of an incompressible fluid in two and three spatial dimensions, and compare our results to direct numerical simulations of the lattice discretized theory.

\end{abstract}

\maketitle

\section{Introduction}

Hydrodynamics is a universal effective theory for the long-time, long-wavelength dynamics of conserved quantities in interacting many-body systems. Based on its universal character, applications in physics range from classical fluids to condensed matter or heavy-ion collisions \cite{landauFluidMechanicsVolume1987,hohenbergTheoryDynamicCritical1977,Jeon:2015dfa}. Hydrodynamics is mathematically described by a classical field theory, in which every space-time point of the system is to be understood as a macroscopic fluid cell much larger than microscopic scales such as the mean free path but much smaller than the scale of macroscopic fluid motion. Non-equilibrium effects manifest themselves in slow variations of the hydrodynamic variables across neighboring fluid cells, leading to dissipative fluxes that enter the hydrodynamic constitutive relations order by order in gradients \cite{landauFluidMechanicsVolume1987,Jeon:2015dfa,Kovtun:2012rj}.

Stochasticity naturally arises from statistical equilibrium fluctuations of individual fluid cells \cite{landauHydrodynamicFluctuations1957,LandauStatMech2,Kapusta:2011gt}. In thermal equilibrium, fluctuations and dissipative effects must balance on average, such that the noise strength and dissipative transport coefficients are not independent quantities but linked via fluctuation-dissipation relations. Importantly, this guarantees that the equilibrium distribution is a stationary distribution of the fluctuating dynamics \cite{taeuber_2014}.

Beyond the requirement of thermodynamic consistency, the inclusion of statistical fluctuations in fluids is known to be important in several respects, ranging from the dissipation range of classical turbulent flows \cite{Bandak_2022} to the dynamics of heavy-ion collisions~\cite{Kapusta:2011gt,Basar:2024srd,Bluhm:2020mpc}. Most prominently, the non-linear interactions between fluctuating modes renormalize transport coefficients such as the shear and bulk viscosities \cite{Andreev:1978,POMEAU197563,Forster:1977zz,Akamatsu:2016llw,Akamatsu:2017rdu,Chattopadhyay:2025uqo}, implying an absolute lower bound on the shear viscosity \cite{Kovtun:2011np,Chafin:2012eq}, and induce non-analytic frequency dependences and long-time tails of hydrodynamic correlation functions \cite{alderDecayVelocityAutocorrelation1970, Forster:1977zz, Kovtun:2011np}. These effects can be computed in a variety of systematic frameworks~\cite{Kovtun:2011np, Forster:1977zz, An:2020vri, An:2019osr,Akamatsu:2017rdu,Akamatsu:2016llw, Harder:2015nxa,Crossley:2015evo,Glorioso:2017fpd,Abbasi:2022aao,Abbasi:2025teu,Gosteva_2026}.

Despite the phenomenological importance of a proper treatment of thermal fluctuations in fluids, the theory of stochastic fluid dynamics encounters fundamental challenges. Since stochastic fluid dynamics is formulated as a classical field theory, it suffers from the same ultraviolet (UV) problems as other classical statistical field theories. While the two-dimensional Navier--Stokes equations with stochastic forcing, as in the mathematical literature, can be given a rigorous meaning without a cutoff \cite{dapratoTwoDimensionalNavierStokes2002}, no comparable result is available for fluctuating hydrodynamics with thermal noise in two or three dimensions. In this physical case, the theory is not expected to have a proper continuum limit and is in practice only well-defined under a suitable regularization scheme \cite{Kovtun:2011np,Basar:2024srd}. For instance, fluctuation corrections to the shear viscosity depend explicitly on the cutoff, growing linearly with it in three and logarithmically in two dimensions \cite{Forster:1977zz,Kovtun:2011np}, such that the bare viscosity is only meaningful in conjunction with a specified regularization.

Still, hydrodynamics provides an effective macroscopic description, and it is physically clear that a hydrodynamic description should not be applicable at arbitrarily small time and distance scales, where the underlying microscopic theory would supplement it as the correct UV completion. Indeed, the local white-noise approximation is expected to break down at short scales, where microscopic calculations or causality arguments yield non-local noise kernels~\cite{Murase:2013tma,Kapusta:2017hfi}. Hence, it seems desirable to introduce a physically motivated cutoff scale into the macroscopic theory of fluid dynamics, to restrict the domain on which the theory is defined to its range of applicability. 

Numerical simulations, which provide one of the most prominent tools for solving the equations of fluctuating hydrodynamics, commonly circumvent this regularization issue by intrinsically tying the regularization of stochastic fluid dynamics to the discretization of the stochastic evolution equations \cite{Florio:2021jlx,Chattopadhyay:2024jlh,Chattopadhyay:2025uqo,Harhoff:2026lty, Basar:2024qxd, Bhambure:2024gnf,Nakano_2026}. Even in this case, fluctuations generally persist on all scales down to the lattice spacing, which then acts as a UV cutoff, thus providing a particular choice of the regularization scheme. Although this scheme can be very practical for numerical simulations, it is rather impractical for \mbox{(semi-)analytic} field-theoretical considerations, which are most straightforwardly performed in the continuum.

However, this choice is not physically well motivated even for numerical simulations, as real-world fluids do not exhibit a discrete structure on macroscopic scales, and the physical cutoff of the effective theory should, as argued above, instead be set by the scale at which the hydrodynamic description breaks down. At the same time, controlling discretization errors requires sending the lattice spacing to zero, a limit in which the strength of the fluctuations diverges. Hence, there is a fundamental conceptual conflict between the role of the lattice spacing as a UV regulator and as the parameter controlling the approach to the continuum limit.

A common remedy is to filter noise above some scale $\ell_\text{filter}$ larger than the lattice spacing \cite{Bluhm:2020mpc}, e.g., by Gaussian smearing, Fourier filtering, or mesh coarsening \cite{Murase:2016rhl, Hirano:2018diu, Singh:2018dpk, Nahrgang:2017oqp, Bluhm:2018plm}. In these approaches, the filter acts on the noise alone while the dissipative terms remain unfiltered, such that fluctuation-dissipation relations are not maintained below $\ell_\text{filter}$, and cutoff-dependent corrections to thermal averages have been observed~\cite{Nahrgang:2017oqp, Bluhm:2018plm}, leaving consistent regularizations an open problem~\cite{Bluhm:2020mpc}.

We therefore propose an alternative approach in which -- inspired by field-theoretical methods -- the classical field theory of stochastic fluid dynamics is first regularized in the continuum and only then discretized for the purpose of numerical simulations. Since, by construction, the regularized theory has a well-defined continuum limit, this also enables direct quantitative comparisons between field-theoretical methods in the continuum and the continuum limit of lattice-discretized numerical simulations. Crucially, an appropriate regularization procedure should not act on noise alone but consistently on all (reversible, dissipative, noise) sectors of the theory, such that important features like symmetries, conservation laws, and fluctuation-dissipation relations survive \cite{Roth:2024rbi,Roth:2024hcu}.

While we believe that this strategy is generally useful, in this paper we explicitly investigate the regularization for an incompressible fluid, and study the renormalization of the shear viscosity in two and three spatial dimensions, extending the classic analysis of Refs.~\cite{Forster:1977zz, Berera:2010rd}. Although we restrict ourselves in this work to non-relativistic hydrodynamics, the same strategy lends itself to relativistic fluctuating hydrodynamics, where recent progress has been made in particular in the density frame \cite{Armas:2020mpr,Basar:2024qxd,Bhambure:2024axa,Bhambure:2024gnf}.

This paper is organized as follows: In Section~\ref{sec:stoch_hydro}, we introduce incompressible stochastic fluids, regularize the theory in a general formalism as described above, and analyze the effect of the regulators at one-loop level. In Section~\ref{sec:RGImprov}, we derive a renormalization group (RG) improved flow equation in order to extract the renormalization of the shear viscosity \mbox{(semi-)analytically} in two and three spatial dimensions. In Section~\ref{sec:numerics}, we discretize the regularized theory while keeping the physical cutoff independent from the lattice spacing, allowing for a continuum limit at fixed physical cutoff. We perform numerical simulations and compare to the \mbox{(semi-)analytical} RG results in Section~\ref{sec:results}, and conclude in Section~\ref{sec:concl_outlook}. Several appendices contain additional details.

\section{Stochastic hydrodynamics} \label{sec:stoch_hydro}
Stochastic hydrodynamics provides a general framework for describing the evolution of thermodynamic fluctuations in fluids. Denoting the hydrodynamic fields as $\phi_a(t,\vec{x})$ with $a$ labeling the fields, we consider the probability density functional $p[\phi_a]$ characterizing the probability to observe the fluid in some state in configuration space. During equilibration, this probability density evolves to a stationary equilibrium $p_\text{eq}$. In a closed system, for example, we have (with $k_B=1$, up to constant factors and conservation constraints)
\begin{align} \label{eq:e_to_S}
    p_\text{eq}[\phi_a] \propto e^{S[\phi_a]},
\end{align}
where the entropy $S$ is the logarithm of the number of microstates \cite{landauStatisticalPhysicsVolume2013}. Instead of solving the Fokker-Planck type evolution equations for $p[\phi_a]$, the Landau-Lifshitz formalism \cite{landauHydrodynamicFluctuations1957, LandauStatMech2} considers dynamic equations for the fluctuating hydrodynamic quantities obtained by promoting the hydrodynamic equations of motion to stochastic partial differential equations. The hydrodynamic conservation laws have the form
\begin{align}\label{eq:general_hydro}
    \partial_t \phi_a + \vec{\nabla}\cdot \vec{J}_a = 0,
\end{align}
in which the flux densities $\vec{J}_a$ comprise contributions
\begin{align}\label{eq:general_hydro_fluxes}
    \vec{J}_a = \vec{J}_a\big|_\text{rev.} + \vec{J}_a\big|_\text{diss.} + \vec{J}_a\big|_\text{fluct.}.
\end{align}
The reversible term $\vec{J}_a|_\text{rev.}$ advects the fluid while conserving entropy, thus keeping the probability measure $p_\text{eq}$ invariant. Conversely, dissipative fluxes $\vec{J}_a|_\text{diss.}$ drive the system towards equilibrium by producing entropy. The third contribution $\vec{J}_a|_\text{fluct.}$ is a stochastic variable containing the effect of fluctuations. Its properties are not arbitrary but fixed to the properties of the dissipative flux $\vec{J}_a\big|_\text{diss.}$ via fluctuation-dissipation relations. The stochastic fluxes render the evolution equations~\eqref{eq:general_hydro} generalized Langevin equations, which encode the evolution of the full probability density functional $p$ in the ensemble of all realizations of the stochastic noise.

\subsection{Incompressible stochastic fluids}
While the regularization procedure we propose in this work is rather general and only relies on a generic formulation of the hydrodynamic equations, we simplify the discussion by considering an incompressible fluid in $d\geq 2$ spatial dimensions. %
In this case, the hydrodynamic equations~\eqref{eq:general_hydro}, \eqref{eq:general_hydro_fluxes} are realized as stochastic Navier--Stokes equations for the momentum density field $\vec{\pi}(t,\vec x)$ (`Model A' by Forster, Nelson and Stephen \cite{Forster:1977zz}),
\begin{align}
    \partial_t \pi_m
    = \mathcal{T}_{mn} \left[-\frac{1}{\rho}\left( \vec{\pi}\cdot\vec{\nabla}\right)\pi_n
    + \frac{\eta}{\rho} \vec{\nabla}^2 \pi_n \right]
    + \xi_m \,, \label{eq:incompr_eom}
\end{align}
with constant mass density $\rho$,
shear viscosity $\eta$ and Gaussian noise $\vec{\xi}(t,\vec x)$ whose moments are fixed by the fluctuation-dissipation theorem (FDT),
\begin{subequations}
\begin{align}
\langle \xi_m(t,\vec{x}) \rangle &= 0 \,, \\
    \langle\xi_m(t,\vec{x})\xi_n(t',\vec{y})\rangle
    &=-2\eta T\,\mathcal{T}_{mn}\vec{\nabla}^2\delta(\vec{x}-\vec{y})\,\delta(t-t') \,.
\end{align}
\end{subequations}
The incompressibility constraint
\begin{align}
    \vec{\nabla}\cdot\vec{\pi}(\vec{x})=0
    \label{eq:incompr}
\end{align}
is ensured by the transverse projector $\mathcal{T}_{mn}$, which acts in Fourier space as\footnote{Alternatively, incompressibility can be realized by replacing the projector with a pressure gradient, with pressure acting as a Lagrange multiplier that adjusts instantaneously to enforce the constraint $\vec{\nabla}\cdot\vec{\pi}(\vec{x})=0$. Solving the resulting Poisson equation and inserting the result is equivalent to the action of $\mathcal{T}_{mn}$ \cite{landauFluidMechanicsVolume1987}.}
\begin{align}
    \mathcal{T}_{mn}(\vec{p})=\delta_{mn}-\frac{p_m p_n}{\vec{p}^2}.
\end{align}

Structurally, in analogy to Eq.~\eqref{eq:general_hydro_fluxes}, the first term on the right-hand side of Eq.~\eqref{eq:incompr_eom} describes the reversible self-advection of the momentum density. The second term describes the dissipation of shear modes in the fluid. The third term describes random shear stress due to thermal fluctuations.

The system relaxes to an equilibrium characterized by the probability density functional
\begin{align}
    p_{\text{eq}}[\vec{\pi}] \propto e^{-H[\vec{\pi}]/T} \delta\left(\vec{\Pi}-\int d^d x \, \vec{\pi}\right), \label{eq:p_eq}
\end{align}
where $T$ denotes the temperature and $\vec{\Pi}$ is the (conserved) value of the total momentum $\int d^d x \,\vec{\pi}(\vec{x})$ fixed by the initial condition. The total kinetic energy 
\begin{align}
    H\left[\vec{\pi}\right] = \int d^d x \, \frac{\vec{\pi}^2(\vec{x})}{2\rho} \label{eq:H_no_R}
\end{align}
serves as an effective Hamiltonian for the distribution.

While the distribution \eqref{eq:p_eq} can be obtained from formally coupling the system to a heat bath of temperature $T$, let us briefly review how it emerges as a limit of the distribution \eqref{eq:e_to_S} discussed in the previous section. Following Ref.~\cite{landauFluidMechanicsVolume1987}, the hydrodynamic variables of an incompressible fluid are the total-energy density $\varepsilon(\vec{x}) = \tilde{\varepsilon}(\vec{x}) + \vec{\pi}^2(\vec{x})/(2\rho)$ (with internal energy $\tilde{\varepsilon}$) and the transverse component $\vec{\pi}(\vec{x})$ of the momentum density. The associated conserved quantities are the total energy $E = \int d^d x \,\varepsilon(\vec{x})$ and the total momentum $\vec\Pi$. Assuming that different fluid cells are statistically independent,\footnote{This assumption has to be weakened near a critical point, since there the correlation length can become comparable to (or even larger than) the size of a fluid cell.} the total entropy of the fluid is given by the integral ${S[\varepsilon,\vec{\pi}] = \int d^dx\,  s(\tilde{\varepsilon}(\vec{x}))}$, where $d^d x\,s(\tilde{\varepsilon}(\vec{x})) $ is the (Boltzmann) entropy of the fluid cell. 
Assuming that the kinetic energy $\vec{\pi}^2/(2\rho)$ is much smaller than the internal energy $\tilde\varepsilon$, we can expand ${s(\tilde{\varepsilon}) \approx s(\varepsilon) - \frac{\vec{\pi}^2}{2\rho T}}$ with temperature defined by $1/T \equiv ds/d\varepsilon$, evaluated at the mean energy density $E/V$. In other words, we neglect the change in temperature caused by the conversion between kinetic energy and internal energy.
Using Eq.~\eqref{eq:e_to_S} and integrating out the total-energy density, we obtain the equilibrium distribution \eqref{eq:p_eq} and conclude that the internal energy, if large, effectively acts as the aforementioned heat bath of temperature $T$.

\subsection{Poisson bracket formalism, symmetries, and conservation laws}
\label{sec:PBs_symm_conservation_laws}
Structurally, the stochastic Navier--Stokes equation \eqref{eq:incompr_eom} represents an example of a dissipative evolution equation of a system with conserved quantities and reversible mode couplings, which in the context of dynamic critical phenomena have been classified in a seminal review by Halperin and Hohenberg \cite{hohenbergTheoryDynamicCritical1977}. Even though it may not be apparent at first sight, the form of the equations of motion~\eqref{eq:incompr_eom} is in fact entirely fixed by general properties like the Hamiltonian~\eqref{eq:H_no_R} along with the symmetries of the system. To clarify these connections, we express the equations of motion \eqref{eq:incompr_eom} in the form \cite{Zwanzig:1961zz,10.1143/PTP.49.764,Kawasaki_1973,PhysRevB.11.4077,hohenbergTheoryDynamicCritical1977,zwanzig2001nonequilibrium,taeuber_2014}  %
\begin{align}
    \partial_t \pi_i (\vec{x}) = \mathcal{T}_{im}\bigg[ &\int_{\vec{y}}\; \{\pi_m (\vec{x}), \pi_n (\vec{y})\} \frac{\delta H[\vec{\pi}]}{\delta \pi_n (\vec{y})} \nonumber\\
    &+ \int_{\vec{y}}\; \Gamma_{mn}(\vec{x}, \vec{y})\frac{\delta H[\vec{\pi}]}{\delta \pi_n(\vec{y})} \bigg] + \xi_i \,,
    \label{eq:eom_general}
\end{align}
with the Poisson bracket \cite{dzyaloshinskiiPoissonBracketsCondensed1980}
\begin{align}
    \{ \pi_m(\vec{x}), \pi_n(\vec{y}) \}
    &=\bigg[ \pi_m(\vec{y})\frac{\partial}{\partial y_n}
    -\pi_n(\vec{x})\frac{\partial}{\partial x_m} \bigg]\delta(\vec{x}-\vec{y})
    \label{eq:bracket}
\end{align}
and the Onsager dissipative kernel
\begin{align}
    \Gamma_{mn}(\vec{x},\vec{y}) = \eta\,\delta_{mn}\,\vec{\nabla}^2_{\vec{x}}\,
    \delta(\vec{x}-\vec{y}).
\end{align}
In this form, it becomes manifest that one has a certain amount of freedom in modifying $\Gamma_{mn}(\vec{x},\vec{y})$ and $H[\vec{\pi}]$ without violating any relevant symmetries. In the following, we discuss the mathematical properties these objects have to satisfy.

\subsubsection{Dissipative terms}

The dissipative kernel $\Gamma_{ij}(\vec{x}, \vec{y})$ is symmetric, as required by Onsager reciprocity, and negative semi-definite, such that under the evolution of the second term in Eq.~\eqref{eq:eom_general},\footnote{\label{fn:projector}The transverse projector is self-adjoint, so that it may be moved from one factor to the other under the spatial integral, and acts as the identity on $\delta H/\delta\pi_n = v_n$, since the fluid velocity is transverse. Hence the integrand reduces to $v_m\Gamma_{mj}v_j$.}
\begin{align}
    \partial_t H\big|_\text{diss.} = \int_{\vec{x},\vec{y}}\frac{\delta H}{\delta \pi_m(\vec{x})} \mathcal{T}_{mi} \Gamma_{ij}(\vec{x},\vec{y}) \frac{\delta H}{\delta \pi_j (\vec{y})} \leq 0.
\end{align}
Negative semi-definiteness renders the dissipative term a gradient flow towards the minimum of $H$, corresponding to entropy production in the general case of Eq.~\eqref{eq:e_to_S}. The noise is fixed in terms of the dissipative kernel, and the FDT in general form reads
\begin{align}
    \langle \xi_m(t,\vec{x})\,\xi_n(t',\vec{y}) \rangle
    = -2T\, \mathcal{T}_{mi} \Gamma_{ij}(\vec{x},\vec{y})
      \mathcal{T}_{jn}\, \delta(t-t'),
      \label{eq:FDT_general}
\end{align}
with the minus sign ensuring positive semi-definiteness of the covariance matrix. Note that the Hamiltonian enters through the thermodynamic force $\delta H / \delta \vec{\pi}$, while the FDT depends only on the dissipative kernel $\Gamma$. Together, these properties make the equilibrium distribution $p_\text{eq}[\vec{\pi}]$ stationary under the dissipative-fluctuating part of the dynamics, as we show by discussing the associated Fokker--Planck equation in Appendix~\ref{sec:fokker_planck}.

The dissipative terms conserve total momentum $\vec{\Pi}$ provided that the dissipative kernel does not couple to the zero mode, $\Gamma_{mn}(\vec{p}=0)=0$. Note, however, that not every dissipative kernel admits a local conservation law in the form of Eq.~\eqref{eq:general_hydro}.

\subsubsection{Reversible terms}
\label{sec:rev_terms}

The Poisson bracket~\eqref{eq:bracket} is fixed by symmetries of the system, as we review here. It can be derived from the transformation behavior of the momentum density under infinitesimal local translations $x'^l = x^l+\delta u^l(\vec{x})$ \cite{dzyaloshinskiiPoissonBracketsCondensed1980}.
In order for the density $\rho$ to remain constant, the coordinate transformation must be volume preserving, $\partial_l \delta u^l(\vec{x}) = 0$. Since the new coordinates $x'^l$ are not necessarily orthonormal, one has to distinguish between co- and contravariant components.\footnote{For volume-preserving transformations, however, one can disregard the distinction between tensors and tensor densities.} For instance, the covariant components $\pi_l(\vec{x})$ of the momentum density transform as
\begin{align}
    \delta \pi_l(\vec{x}) = -\delta u^m(\vec{x})\frac{\partial}{\partial x^m} \pi_l(\vec{x}) - \pi_m(\vec{x}) \frac{\partial}{\partial x^l} \delta u^m(\vec{x}) \,.
\end{align}
One can easily verify that this can be written as
\begin{align}
    \delta \pi_l(\vec{x}) = \int_{\vec{y}} \{ \pi_l(\vec{x}), \pi_m(\vec{y}) \} \delta u^m(\vec{y}),  \label{eq:pi_transform_PB}
\end{align}
or, equivalently,
\begin{align}
\frac{\delta \pi_l(\vec{x})}{\delta u^m(\vec{y})} =
    \{ \pi_l(\vec{x}), \pi_m(\vec{y}) \}, \label{eq:def_PB}
\end{align}
which can be regarded as the definition of the Poisson bracket~\eqref{eq:bracket}. Since we work in Cartesian coordinates and all transformations considered are volume preserving, we shall not distinguish between upper and lower indices in the following. Defining the fluid velocity
\begin{align}
    v_m(\vec{x}) \equiv \frac{\delta H[\vec{\pi}]}{\delta \pi_m(\vec{x})} \,, \label{eq:fluid_vel}
\end{align}
a comparison of the reversible term in Eq.~\eqref{eq:eom_general} with Eq.~\eqref{eq:pi_transform_PB} shows that, during a time step $dt$, the reversible dynamics acts on the momentum density as a local translation by $\delta u_l = v_l\,dt$, i.e., as (self-)advection of the fluid. Since $\vec{v}$ is transverse, this translation is indeed volume preserving. Explicitly, the reversible term assumes the form
\begin{align}
\begin{split}
    \partial_t \pi_m(\vec{x})\big|_\text{rev.}
    = -\int_{\vec{y}} &\mathcal{T}_{mn}(\vec{x}-\vec{y})\, v_l(\vec{y})\\
    &\times\left(\frac{\partial \pi_n(\vec{y})}{\partial y_l} - \frac{\partial \pi_l(\vec{y})}{\partial y_n}\right) \,.
\end{split}
\label{eq:rev_eom}
\end{align}
In particular, for $\vec{v}=\vec{\pi}/\rho$ (i.e.,~$H=\int \vec{\pi}^2/(2\rho)$), the second term is a total gradient $v_l\partial_n\pi_l = \partial_n[\vec{\pi}^2/(2\rho)]$ which vanishes upon transverse projection, leaving the familiar self-advection term in the stochastic Navier--Stokes equation \eqref{eq:incompr_eom}.

One particular advantage of the Poisson bracket formulation is that it makes symmetries (and related conservation laws) manifest. We begin with the Hamiltonian. Conservation of $H[\vec{\pi}]$ is a consequence of the antisymmetry of the Poisson bracket,
\begin{align}
    \partial_t H\big|_\text{rev.}
    &= \int_{\vec{x}} \frac{\delta H[\vec{\pi}]}{\delta \pi_i(\vec{x})}\,
       \partial_t \pi_i(\vec{x})\big|_\text{rev.} \nonumber \\  
    &= -\int_{\vec{x}\vec{y}}
       v_m(\vec{x})\mathcal{T}_{mn}(\vec{x}-\vec{y})\, v_l(\vec{y})\left(\frac{\partial \pi_n(\vec{y})}{\partial y_l} - \frac{\partial \pi_l(\vec{y})}{\partial y_n}\right) \nonumber\\
       &=0,
\end{align}
where the transverse projector has been removed using the transversality of $v_m(\vec{x})$ (cf.\ footnote~\ref{fn:projector}). The remaining integrand contracts an antisymmetric kernel with a symmetric product and thus vanishes. %
However, conservation of $H[\vec{\pi}]$ is not yet sufficient to guarantee that~\eqref{eq:p_eq} is the equilibrium distribution. For this to hold, the probability density current generated by the reversible step -- regarded as a vector field on the (phase) space of all $\vec{\pi}(\vec{x})$ -- must be divergence-free (see, e.g.,~Ref.~\cite{taeuber_2014}), 
\begin{align}
    \int_{\vec{x}} \frac{\delta}{\delta \pi_l(\vec{x})} \left(  p_{\text{eq}}[\vec{\pi}] \partial_t \pi_l(\vec{x}) \big|_\text{rev.} \right) = 0, \label{eq:divFreeCond}
\end{align}
which can be seen from the corresponding Fokker--Planck equation. Here this is indeed the case, as outlined in Appendix~\ref{sec:fokker_planck}.

Conservation of the total momentum $\vec{\Pi}$ on the other hand is implied by translational invariance of $H[\vec{\pi}]$, as expected from Noether's theorem. 
The chain rule entails that the change in the Hamiltonian $\delta H[\vec{\pi}]$ under an infinitesimal global translation $\delta u_m(\vec{x})=\text{const.}$ inherited from \eqref{eq:pi_transform_PB} is given by
\begin{align}
        \delta H &= \left( \int_{\vec{x}\vec{y}} v_l(\vec{x}) \{ \pi_l(\vec{x}), \pi_m(\vec{y}) \} \right) \delta u_m \,.
\end{align}
In a translationally invariant system $\delta H=0$ for all directions $m$, which entails that the term in parentheses must vanish. The antisymmetry of the Poisson bracket then immediately implies that the total momentum $\Pi_m$ is conserved under reversible dynamics,
\begin{align}
    \partial_t \Pi_l\big|_\text{rev.} &= \int_{\vec{x}\vec{y}\vec{z}} \mathcal{T}_{lm}(\vec{x}-\vec{y}) \{\pi_m(\vec{y}),\pi_n(\vec{z})\} v_n(\vec{z}) \nonumber \\
    &= -\int_{\vec{y}\vec{z}} v_n(\vec{z}) \{\pi_n(\vec{z}),\pi_m(\vec{y})\}   \int_{\vec{x}} \mathcal{T}_{lm}(\vec{x}-\vec{y}) \nonumber \\
    &= 0, \label{eq:Pi_conserv}
\end{align}
where in the second line the integral over $\vec{x}$ simply yields the transverse projector of the zero mode, which depends on convention but is generally a multiple of $\delta_{lm}$. Note, however, that in lattice-discretized theories momentum conservation has to be verified explicitly, since the group of spatial translations is discrete and this argument is thus not directly applicable.

Finally, one expects invariance of the theory under boosts by a constant velocity $\vec{V}$. Under the coordinate transformation $\vec{x}' = \vec{x}+\vec{V}t$, we demand that the fluid velocity \eqref{eq:fluid_vel} transforms as
\begin{align}
    \vec{v}'(t,\vec{x}') = \vec{v}(t,\vec{x}) +\vec{V} \,.
\end{align}
For a quadratic Hamiltonian of the form
\begin{align}
    H[\vec{\pi}] = \int_{{\vec{x}\vec{y}}} \frac{1}{2} \vec\pi(\vec x) M^{-1}(\vec x-\vec y)\vec\pi(\vec y), \label{eq:H_quadr}
\end{align}
the relation \eqref{eq:fluid_vel} implies that the momentum density has to transform as
\begin{align}
    \vec{\pi}'(t,\vec{y}') = \vec{\pi}(t,\vec{y}) + \left(\int_{{\vec z'}} M(\vec y'-\vec z') \right) \vec{V} \,. \label{eq:pi_Galilean_trafo}
\end{align}
Indeed, it is straightforward to verify that if $\vec\pi(t,\vec x)$ solves the equations of motion \eqref{eq:eom_general}, then so does $\vec\pi'(t,\vec x')$, expressing Galilean invariance of the theory.

\subsection{Regularization}
\label{sec:reg}

We have now introduced a formalism that allows us to deform the theory while keeping the relevant symmetries manifest. To this end, we will in this section introduce UV cutoffs $\Lambda$ into the definitions of the Hamiltonian $H$ and dissipative kernel $\Gamma$ that reduce to the previous definitions in the limit $\Lambda \to \infty$. We leave the Poisson bracket untouched, as it is fixed by the transformation behavior of $\vec{\pi}$ under local translations.

Defining $\int_{\vec{p}}\equiv \int \frac{d^d p}{(2\pi)^d}$, we promote the Hamiltonian~\eqref{eq:H_no_R} in Fourier space to
\begin{align}
    H_\Lambda[\vec{\pi}]&=\int_{\vec{p}} \vec{\pi}(-\vec{p})\frac{ 1+R_\Lambda(\vec{p})}{2\rho}\vec{\pi}(\vec{p}).%
    \label{eq:h_reg}
\end{align}
Assuming the regulator to be real and furthermore dimensionless and isotropic, we may write $R_\Lambda(\vec{p})=r(p/\Lambda)$ with $p=|\vec{p}|$. Note that $R_\Lambda$ acts on the static properties of the theory as it modifies the equilibrium distribution $p_\text{eq}$ and thereby the thermal spectrum, while leaving the FDT untouched. Since high energy amounts to an exponential suppression of modes in $p_\text{eq}$, we demand $r(x)\geq 0$ everywhere and $r(x)\gg 1$ in the UV $x\to\infty$. Moreover, we require $r(0)=0$, such that a boost by $\vec{V}$ shifts the momentum density by $\rho\vec{V}$ [cf.~Eq.~\eqref{eq:pi_Galilean_trafo}] and the long-wavelength properties of the theory remain unaffected. We will later see that this is a comparatively weak deformation, since for an incompressible fluid $H$ is quadratic anyway, such that there are no static couplings to be renormalized.  However, the regulator has an indirect effect on the dynamics of the fluid via its effective velocity
\begin{align}
    v_i(\vec{y})=\frac{\delta H_\Lambda}{\delta \pi_i(\vec{y})}=\frac{1}{\rho}\left(\pi_i(\vec{y})+\int_{\vec{z}} R_\Lambda(\vec{y}-\vec{z}) \pi_i(\vec{z})\right). \label{eq:fluid_vel_reg}
\end{align}
Unlike in the unregularized theory, the second term of the reversible equation~\eqref{eq:rev_eom} is no longer a pure gradient and hence survives the transverse projection.

Secondly, we can introduce a UV regulator by redefining the dissipative kernel as
\begin{align}
    \Gamma_\Lambda(\vec{x}-\vec{y})
    =
    \eta (\delta(\vec{x}-\vec{y})+K_\Lambda(\vec{x}-\vec{y}))\vec{\nabla}_{\vec{x}}^2, \label{eq:reg_kin_coeff}
\end{align}
with the regulator $K_\Lambda$ having the same properties as $R_\Lambda$. In particular, we require $K_\Lambda(\vec{p}=0)=0$, such that the regulator leaves the long-wavelength dynamics unaffected and $\eta$ retains its meaning as the shear viscosity on large scales. Momentum conservation, on the other hand, imposes no further constraint, since the Laplacian in Eq.~\eqref{eq:reg_kin_coeff} already ensures $\Gamma_\Lambda(\vec{p}=0)=0$. The dissipative term in the equations of motion~\eqref{eq:eom_general} then becomes
\begin{align}
    \partial_t \pi_i\big|_\text{diss.}
    &= \int_{\vec{y}}\; \Gamma_\Lambda(\vec{x}-\vec{y})\,
       \frac{\delta H_\Lambda}{\delta \pi_i(\vec{y})} \\ \nonumber 
    &= \eta \left( (\vec{\nabla}^2 v_i)(\vec{x})
       + \int_{\vec{y}} K_\Lambda(\vec{x}-\vec{y})\,
         (\vec{\nabla}^2 v_i)(\vec{y}) \right),
\end{align}
and the FDT immediately fixes the noise correlator to
\begin{align}
\begin{split}
    \langle\xi_m(t,\vec{p})\xi_n(t',\vec{p}')\rangle
    =&\; 2\eta T \vec{p}^2\left[1+K_\Lambda(\vec{p})\right]\mathcal{T}_{mn}(\vec{p}) \\
      &\times(2\pi)^d\delta(\vec{p}+\vec{p}')\delta(t-t'),
\end{split} \label{eq:noise_reg}
\end{align}
such that the introduction of the regulator $K_{\Lambda}$ effectively amounts to a scale-dependent shear viscosity $\eta[1+K_\Lambda(p)]$. %
With these regularizations it is then straightforward to derive the modified form of the stochastic evolution equations from Eq.~\eqref{eq:eom_general}, yielding
\begin{align}
    \partial_t \pi_l(\vec{x}) = \mathcal{T}_{lm}\bigg[ &-\, v_n(\vec{x}) \left(\frac{\partial \pi_m(\vec{x})}{\partial x_n} - \frac{\partial \pi_n(\vec{x})}{\partial x_m}\right) \nonumber \\
    &+ \int_{\vec{y}} \Gamma_\Lambda(\vec{x}-\vec{y})\, v_m(\vec{y}) \bigg] + \xi_l \,, \label{eq:reg_eom}
\end{align}
with the regulated fluid velocity~\eqref{eq:fluid_vel_reg}, the regulated dissipative kernel~\eqref{eq:reg_kin_coeff}, and the noise correlator~\eqref{eq:noise_reg}.

We finally note that, as explained in the previous section, the structure of the Poisson bracket conserves our new choice of $H_\Lambda$ under the reversible part of the dynamics. The dissipative kernel is symmetric, given that $K_\Lambda$ is real and isotropic. Since $1+K_\Lambda\geq 0$, it is also negative semi-definite. Translational invariance of $H_\Lambda$ due to $R_\Lambda(\vec{x},\vec{y})=R_\Lambda(\vec{x}-\vec{y})$ implies momentum conservation, as discussed previously, and $H_\Lambda$ being of the form \eqref{eq:H_quadr} (with $M^{-1} = (1+R_{\Lambda})/\rho$) guarantees Galilean invariance. Hence, all relevant symmetries are preserved.

We conclude this section by remarking that the two regulators play different roles: $R_\Lambda(q)$ suppresses static fluctuations of UV modes by increasing their energy, whereas $K_\Lambda(q)$ leaves all static properties of the system invariant but causes a faster relaxation of UV modes. In the case of our incompressible fluid the static fluctuations of $\vec{\pi}$ are Gaussian and thus do not need to be regulated. As such, $K_\Lambda(q)$ suffices to render the shear viscosity UV finite (at least at the one-loop level), as we shall see in the next section. %
Nevertheless, in a general setting, %
we expect that both types of regulators will be required to regularize the divergences of the classical-statistical field theory. Moreover, even in the absence of divergences, a static regulator may also be useful for the numerical discretization.

\subsection{Effect of the regulators}
\label{sec:effect_of_regulators}

Now in order to understand the effect of the different regulators regarding the regularization of the theory, it proves insightful to study the renormalization of the theory perturbatively. Diagrammatically, the one-loop correction to the (retarded) self-energy $\Sigma^{\tilde{\pi}_l \pi_m}(\omega,\vec{p})$ is  (cf.~Appendix~\ref{sec:one_loop_calc})
\begin{align}
    \Sigma^{\tilde{\pi}_l \pi_m}(\omega,\vec{p}) &= 
    i\;\; \begin{tikzpicture}[baseline=-0.5ex]
        \draw[] plot[domain=0:180, samples=200, smooth, variable=\a] ({0.5*cos(\a)*(1.0-0.1*sin(12*\a))},{0.5*sin(\a)*(1.0-0.1*sin(12*\a))});
        \draw[] plot[domain=180:270, samples=200, smooth, variable=\a] ({0.5*cos(\a)*(1.0-0.1*sin(12*\a))},{0.5*sin(\a)*(1.0-0.1*sin(12*\a))});
        \draw[densely dotted] plot[domain=270:360, samples=200, smooth, variable=\a] ({0.5*cos(\a)*(1.0-0.1*sin(12*\a))},{0.5*sin(\a)*(1.0-0.1*sin(12*\a))});
		\draw[densely dotted,smooth,domain=0:0.5,variable=\x] plot (-\x-0.5,{-0.05*sin(8*3.14159*\x r)});
        \draw[smooth,domain=0:0.5,variable=\x] plot (\x+0.5,{0.05*sin(8*3.14159*\x r)});
	 	\fill[black] (-0.5,0) circle (0.05);
	 	\fill[black] (+0.5,0) circle (0.05);
    \end{tikzpicture}. \label{eq:self_en_corr}
\end{align}
The correction to the shear viscosity $\Delta\eta$ can be read off from the self-energy at small frequencies and small wavenumbers %
(see, e.g., Ref.~\cite{Kovtun:2012rj}), which is expected to be 
\begin{align}
    \Sigma^{\tilde{\pi}_l \pi_m}(\omega,\vec{p}) =  - \frac{\Delta\eta}{\rho} (1+K_{\Lambda}(p))(1+R_{\Lambda}(p)) \vec{p}^2 \mathcal{T}_{lm}(\vec{p}) + \cdots \label{eq:eta_corr_from_self_en}
\end{align}
A straightforward but lengthy calculation yields the following expression for $\Delta\eta$ (cf.~Appendix~\ref{sec:one_loop_calc}) 
\begin{align}
    \Delta \eta &=  \frac{S_{d-1}   g^2 \rho  T}{(2\pi)^d 2d (d+2) \eta}\int\displaylimits_{0}^{\infty}  \frac{dq \, q^{d-3}  }{  \left(K_{\Lambda }\left(q\right)+1\right) \left(R_{\Lambda }\left(q\right)+1\right)^3} \times \nonumber \\
    &\bigg\{\left(d^2-2\right) \left(R_{\Lambda }\left(q\right)+1\right){}^2+2 (d-1) q^4 \left(\frac{\partial R_{\Lambda }\left(q\right)}{\partial q^2}\right)^2 \nonumber \\
    &\hspace{1.0cm} -4 q^2 \left(R_{\Lambda }\left(q\right)+1\right) \frac{\partial R_{\Lambda }\left(q\right)}{\partial q^2} \bigg\} \label{eq:Del_eta}
\end{align}
for $d>2$, where $S_{d-1} = 2\pi^{d/2}/\Gamma(d/2)$ denotes the surface area of the unit $(d-1)$-sphere. Moreover, we have introduced the mode-coupling constant $g$ in front of the Poisson bracket term in the equations of motion \eqref{eq:eom_general} to serve as the expansion parameter of perturbation theory; it will be set back to $g=1$ in our results below.

While the above expression encapsulates the effect of both UV regulators, we will consider them separately in the following to isolate their effects.
When only the regulator $R_{\Lambda}(q)$ in the Hamiltonian is present and the viscosity regulator $K_{\Lambda}(q) = 0$ is set to zero, one finds that the $q\to\infty$ contribution to the integral behaves as 
\begin{align}
    \Delta \eta &\sim \int\displaylimits^{\infty}  \frac{dq \, q^{d-3}  }{R_{\Lambda }\left(q\right)} \, \label{eq:Del_eta_R}
\end{align}
for large values of $q$. Hence, the one-loop correction is indeed UV finite if $R_{\Lambda}(q)$ grows faster than $q^{d-2}$ for $q\to\infty$. However, it is not possible to recover the sharp cutoff that is usually used in perturbative computations \cite{Forster:1977zz,Berera:2010rd,Kovtun:2011np,Chafin:2012eq} with this regularizing procedure. To see this, consider for instance the following family of regulators,
\begin{align}
    R_{\Lambda}(q) = \left(\frac{q^2}{\Lambda^2}\right)^{\alpha} \,, \quad \alpha > (d-2)/2 \,,
\end{align}
for which the result for the one-loop correction \eqref{eq:Del_eta} is
\begin{align}
    \Delta \eta &= \frac{\widetilde{c}_d g^2 \rho  T \Lambda^{d-2}}{\eta}
\end{align}
with
\begin{align}
    \widetilde{c}_d &= \frac{\Omega_d}{(2\pi)^d} \frac{(4+(d-1)(2\alpha(d-2)+d^2))\pi}{16\alpha(d+2) \sin((d-2)\pi/2\alpha)},
\end{align}
where $\Omega_d \equiv S_{d-1}/d$ denotes the volume of the unit $d$-ball. 
The limit $\alpha \to \infty$ realizes a sharp cutoff in the equilibrium distribution \eqref{eq:p_eq}: modes with $p > \Lambda$ are ``infinitely'' suppressed, while modes with $p<\Lambda$ are not regulated at all.
However, in this limit the coefficient and thus the one-loop correction \eqref{eq:Del_eta} diverges,
\begin{align}
    \widetilde{c}_d \to \frac{\Omega_d}{(2\pi)^d} \frac{(d-1)\alpha}{4(2+d)} \,, \quad \alpha\to\infty \,,
\end{align}
which comes from the region $q \approx \Lambda$ in the integral \eqref{eq:Del_eta}.

On the other hand, when we set the Hamiltonian regulator $R_{\Lambda}(q) = 0$ to zero and keep the viscosity regulator $K_{\Lambda}(q)$,
we are left with
\begin{align}
    \Delta \eta &=  \frac{S_{d-1}  (d^2-2) g^2 \rho  T}{(2\pi)^d 2d (d+2)   \eta} \int\displaylimits_{0}^{\infty}  \frac{dq \, q^{d-3}}{ K_{\Lambda }\left(q\right)+1}. \label{eq:one_loop_corr_int}
\end{align}
We observe that the integral is indeed UV finite if $K_{\Lambda}(q)$ diverges faster than $q^{d-2}$ for $q \to \infty$.  Writing $K_{\Lambda}(q) = k(q/\Lambda)$ and substituting $q = \Lambda z$, the one-loop correction~\eqref{eq:one_loop_corr_int} can be written in the convenient form
\begin{align}
    \Delta \eta = \frac{c_d \rho g^2 T \Lambda^{d-2}}{\eta} \label{eq:eta_corr_one_loop}
\end{align}
for $d>2$, with the dimensionless coefficient
\begin{align}
    c_d &= \frac{S_{d-1}(d^2-2)}{(2\pi)^d 2 d (d+2) } \int\displaylimits_{0}^{\infty}  \frac{dz\,z^{d-3}}{ k\left(z\right)+1} \label{eq:cd} .
\end{align}
Specifically, for the family of regulators
\begin{align}
    k_{\alpha}(z) = (z^{2\alpha}-1) \theta(z^2-1) \,, \quad \alpha>(d-2)/2 \,, \label{eq:reg_family}
\end{align}
the coefficient is
\begin{align}
    c_d = \frac{\Omega_d(d^2-2)\alpha}{(2\pi)^d (d^2-4)(2+2\alpha-d)}.
\end{align}
In the limit $\alpha\to \infty$ the regulator~\eqref{eq:reg_family} corresponds to a sharp cutoff, for which we recover the result $c_{3} \to 7/60\pi^2$ for a sharp UV cutoff\footnote{Note that introducing a sharp UV cutoff \emph{before} evaluating the wavevector integrals requires careful treatment of the wavevector routing through the diagram \cite{Berera:2010rd}.} originally derived by Forster, Nelson and Stephen \cite{Forster:1977zz} (see also \cite{Kovtun:2011np,Chafin:2012eq}).

So far we have only discussed the behavior for $d>2$, where the perturbative result is free of infrared divergencies. However, for two-dimensional fluids ($d=2$) the integral \eqref{eq:cd} is logarithmically infrared divergent, while the UV divergence has readily been eliminated by virtue of the UV regulator. By introducing a sharp infrared cutoff $\mu$ to the $q$-integral as an additional infrared regulator in \eqref{eq:one_loop_corr_int}, we find that the result in \eqref{eq:eta_corr_one_loop} remains valid, however with the coefficient
\begin{align}
    c_{2} &= \frac{1}{16\pi} \int\displaylimits_{\mu/\Lambda}^{\infty}  \frac{dz}{\left( k\left(z\right)+1 \right) z}
    = \frac{1}{16\pi} \ln e^{\frac{1}{2\alpha}} \frac{\Lambda}{\mu}  \label{eq:c2}
\end{align}
experiencing a logarithmic dependence on the infrared cut-off $\mu$. While it is clear that the details of the infrared and UV regularization will affect the scale coefficient $e^{\frac{1}{2\alpha}}$ of the logarithm, we emphasize that the prefactor $1/(16\pi)$ in front of the logarithm is universal.

\section{RG improved flow equation}
\label{sec:RGImprov}

From the one-loop expression~\eqref{eq:Del_eta} we can easily derive a renormalization group (RG) improved expression for the renormalized shear viscosity inspired by the real-time functional renormalization group (FRG) framework developed in \cite{Roth:2024rbi,Roth:2024hcu}. Since the regularization discussed above is to some extent inspired by the real-time FRG, it turns out that to derive the RG improvement, we can proceed along similar lines.

In the framework developed in \cite{Roth:2024rbi,Roth:2024hcu}, the real-time FRG flow is formulated by introducing a smooth infrared (IR) regulator to the effective Hamiltonian~\eqref{eq:H_no_R} chosen so as to suppress fluctuations from modes with wavenumbers $\vec{p}^2 \lesssim \kk^2$, where $\kk$ is the FRG scale. Introducing the regulator in this way ensures that the symmetries discussed in Sec.~\ref{sec:PBs_symm_conservation_laws} are preserved during the flow \cite{Roth:2024rbi,Roth:2024hcu}, and by lowering the FRG scale from $\kk \gg \Lambda$ to $\kk\to 0$, we successively integrate out statistical fluctuations.

To derive the RG improvement, we adapt this procedure by simply adding an additional IR regulator $R_{\kk}$ to the regulator $R_{\Lambda}$ from Sec.~\ref{sec:reg}, i.e., we simply shift $ R_{\Lambda}\to R_{\Lambda}+R_{\kk}$ in the regulated expression for the one-loop correction~\eqref{eq:Del_eta}. Next, note that the  resulting equation for the $R_{\kk}$-dependent one loop correction $\Delta \eta_{\kk}$ has the structure 
\begin{align}
    \Delta \eta_{\kk} = \frac{g^2 \rho T}{\eta} \int\displaylimits_{0}^{\infty} dq\,f_{\kappa\Lambda}(q)
\end{align}
where the integrand $f_{\kappa\Lambda}(q)$ depends on the regulators $R_\kk$, $R_\Lambda$ and $K_\Lambda$, but is independent of $\rho$, $g$, $T$ and $\eta$.

Differentiating with respect to $\kk$ and  replacing the bare viscosity $\eta$ with the scale ($\kk$) dependent $\eta_{\kk}$ on the right-hand side yields the RG improved flow equation
\begin{align}
    \partial_{\kk} \eta_{\kk} = \frac{g^2 \rho T}{\eta_{\kk}} \int\displaylimits_{0}^{\infty} dq\,\partial_{\kk} f_{\kappa\Lambda}(q) \,. \label{eq:eta_flow}
\end{align}

The initial condition for the RG flow is set by the bare viscosity, $\eta_{\kk_{\text{UV}}} \approx \eta$, since for $\kk_{\text{UV}}\gg\Lambda$ all fluctuations are effectively suppressed due to the presence of both the IR and a UV regulator. On the other hand, at $\kk=0$ all fluctuations have been integrated out, such that one obtains the fully renormalized viscosity, $\eta_{\kk=0} = \eta_R$.

Eq.~\eqref{eq:eta_flow} can be solved analytically. By multiplying both sides with $\eta_\kk$, the left-hand side becomes a total derivative, $\tfrac{1}{2}\partial_{\kk} \eta_{\kk}^2$. Integrating from $\kk=0$ to $\kk = \kk_{\text{UV}} \to \infty$ yields 
\begin{align}
    &\frac{1}{2} (\eta_{R}^2-\eta^2) =\rho g^2 T \int\displaylimits_{0}^{\infty} dq  (f_{\kappa=0,\Lambda}(q)-f_{\kappa=\infty,\Lambda}(q)).
\end{align}

In particular, $f_{\kappa=\infty,\Lambda}(q)=0$ due to the divergent IR regulator $R_{\kk}$, such that the integral on the right-hand side has the same form as the one in the one-loop expression~\eqref{eq:one_loop_corr_int}. Re-arranging, we obtain
\begin{align}
    \eta_R = \sqrt{\eta^2+ 2c_d \rho g^2 T \Lambda^{d-2}}, \label{eq:etaR_result}
\end{align}
where $c_d$ is the same dimensionless coefficient as in the one-loop correction~\eqref{eq:cd}. For large $\eta \gg \sqrt{2c_d \rho g^2 T \Lambda^{d-2}}$, we may expand the square root,
\begin{align}
    \eta_R \approx \eta + \frac{c_d \rho g^2 T \Lambda^{d-2}}{\eta} + \cdots \,,
\end{align}
recovering the one-loop result~\eqref{eq:eta_corr_one_loop}.
Beyond the one-loop result, however, we find that $\eta_R$ is a monotonic function of $\eta$, and approaches the value $\sqrt{2c_d \rho g^2 T \Lambda^{d-2}}$ in the inviscid limit $\eta\to 0$. We finally note that the general form of the result~\eqref{eq:etaR_result} is independent of the specific choice of the IR regulator $R_{\kk}(p)$, it is however sensitive to the UV regularization of the theory, which enters through the coefficient $c_d$ (see~\eqref{eq:cd}).\\

While the above discussion is framed for $d>2$, the re-summed result \eqref{eq:etaR_result} can be straightforwardly extended to $d=2$ by regularizing the (logarithmically) IR divergent integral in \eqref{eq:eta_flow} with a sharp IR cutoff $\mu$. In this way \eqref{eq:etaR_result} remains valid also for two-dimensional fluids with the coefficient $c_2$ determined by Eq.~\eqref{eq:c2}, i.e.,
\begin{align}
    \eta_R = \sqrt{\eta^2+ \frac{\rho g^2 T}{8\pi} \ln e^{\frac{1}{2\alpha}} \frac{\Lambda}{\mu} } \label{eq:etaR_result_2d}
\end{align}
In particular, in the limit of a sharp UV cutoff ($\alpha\to \infty$) this recovers the classic result by Forster, Nelson and Stephen \cite{Forster:1977zz}.

\subsection{Self-consistent wavenumber dependence}
\label{sec:self_cons_RG_improv}

By including the full wavenumber ($p$) dependence of the one loop perturbative result, the idea of RG improvement can be generalized to derive an RG improved flow equation for the full wavenumber ($p$) dependence of the shear viscosity $\eta_{\kappa}(p)$. By generalizing \eqref{eq:eta_corr_from_self_en} to finite $\vec{p}$ we can define a $p$-dependent correction to the shear viscosity $\Delta \eta_{\kappa}(p)$,
\begin{align}
    \Sigma_{\kappa}^{\tilde{\pi}_l \pi_m}(0,\vec{p}) =  - \frac{\Delta\eta_{\kappa}(p)}{\rho} (1+K_{\Lambda}(p)) (1+R_{\kappa}(p)) p^2 \mathcal{T}_{lm}(\vec{p})  \label{eq:eta_corr_from_self_en_p}
\end{align}
In practice, it turns out to be more convenient to work instead with the correction to the dissipative kernel 
$\Delta\Gamma_{\kappa}(p) = -\Delta\eta_{\kappa}(p)(1+K_{\Lambda}(p)) p^2$, which is then related to the self energy via
\begin{align}
    \Delta \Gamma_{\kk}(p) = \frac{\mathcal{T}_{lm}(\vec{p})}{d-1} \frac{\rho \Sigma_\kk^{\tilde{\pi}_l\pi_m}(0,\vec{p})}{1+R_{\kk}(p)} \label{eq:kin_coeff_self_en},
\end{align}
where the vector $\vec{p}$ on the right-hand side can have an arbitrary direction with fixed magnitude $|\vec{p}| = p$ due to spatial isotropy. Differentiating~\eqref{eq:kin_coeff_self_en} with respect to $\kk$ and replacing the bare dissipative kernel $\Gamma$ with the renormalized dissipative kernel $\Gamma_\kk$ on the right-hand side 
\begin{align}
    \partial_\kk \Gamma_\kk(p) = \partial_\kk \left\{ \frac{\mathcal{T}_{lm}(\vec{p})}{d-1} \frac{\rho \Sigma_\kk^{\tilde{\pi}_l\pi_m}(\omega=0,\vec{p})}{1+R_{\kk}(p)} \right\} \bigg\rvert_{\Gamma \to \Gamma_\kk}
\end{align}
yields an RG improved flow equation for $\Gamma_\kk(p)$ (using the $\vec{p}$-dependent one-loop correction to the self energy that is provided in Eq.~\eqref{eq:p_dep_self_en} in the appendix).

In practice, we solve the flow equation numerically with the methods outlined in Appendix~\ref{sec:rFRG_num_impl}. The initial condition is given by $\Gamma_{\kk_\text{UV}}(p) = -\eta p^2(1+K_{\Lambda}(p))$ at a sufficiently large UV scale $\kk_{\text{UV}} \gg \Lambda$, and we employ a smooth exponential (IR) regulator
\begin{equation}
    R_{\kk}(p) = \frac{p^2/\Lambda^2}{e^{p^2/\kk^2}-1} \,,
\end{equation}
where we have introduced a factor $\Lambda^2$ to render $R_{\kk}(p)$ dimensionless. Since in $d=2$ spatial dimensions the correction to the self energy is logarithmically IR divergent, we regulate this IR divergence in our numerical setup by introducing a sharp IR cutoff $\mu$. We note that since a finite external wavenumber $\vec{p} \neq 0$ itself acts as an IR cutoff, the resulting shear viscosity $\eta_R(p)$ is independent of $\mu$ for $p \gg \mu$, which allows us to define a $\mu$-independent $\eta_R(p)$ by taking the limit $\mu\to 0$. In this case, the characteristic logarithmic divergence of $\eta_R$ (cf.~Eqs.~\eqref{eq:c2}, \eqref{eq:etaR_result}) is reflected in the $p\to 0$ limit of $\eta_R(p)$, as we shall see in Sec.~\ref{sec:results_2d} below.

\section{Numerical simulations of the regularized theory}
\label{sec:numerics}

In this section, we proceed by discretizing the regularized equations of motion~\eqref{eq:reg_eom} on a spatial lattice, making them amenable to numerical simulations. We do so while preserving the structures discussed in Sec.~\ref{sec:PBs_symm_conservation_laws}, by discretizing only the objects that define the dynamics instead of the equations of motion term by term. Crucially, we keep the cutoff $\Lambda$ and the lattice spacing $a$ as independent parameters, allowing for a continuum limit $a\to 0$ at fixed physical value of the cutoff $\Lambda$.

Working on a $d$-dimensional cubic lattice with periodic boundary conditions, $N^d$ sites, and spacing $a$, we replace 
\begin{align}
    \int_{\vec{y}}&\to a^d \sum_{\vec{y}},
    \qquad
    \frac{\delta}{\delta \pi_n(\vec{x})} \to a^{-d} \frac{\partial}{\partial \pi_n(\vec{x})}.
\end{align}
We denote the box length as $L=Na$ and volume as $V=L^d$. Lattice fields $f(\vec x)$ are Fourier-transformed as
\begin{subequations}
\begin{align}
    f(\vec p) &= a^d \sum_{\vec x} e^{-i\vec p\cdot\vec x} f(\vec x), \\
    f(\vec x) &= \frac{1}{V} \sum_{\vec p} e^{i\vec p\cdot\vec x} f(\vec p),
\end{align}
\end{subequations}
with lattice sites and wavenumbers
\begin{subequations}
\begin{align}
    x_m &= a\, j_m, && j_m \in \{0,\dots,N-1\}, \\
    p_m &= \frac{2\pi n_m}{L}, && n_m \in \{-N/2+1,\dots,N/2\},
\end{align}
\end{subequations}
for even $N$. Although we set the regulator $R_\Lambda=0$ in all simulations, we keep it here to emphasize that the construction does not depend on this choice.

Discretizing the Hamiltonian~\eqref{eq:h_reg}, we define
\begin{subequations}
\begin{align}
    H_\Lambda
    &= \frac{a^{2d}}{2\rho}\sum_{\vec x,\vec y}\pi_n(\vec x) \left[\frac{\delta_{\vec x\vec y}}{a^d}+R_\Lambda(\vec x-\vec y)\right]\pi_n(\vec y)\\
    &=\frac{1}{2\rho V}\sum_{\vec p}\big[1+R_\Lambda(\vec p)\big]\,|\vec\pi(\vec p)|^2.
\end{align}
\end{subequations}
Since $R_\Lambda(\vec x-\vec y)$ is real and symmetric, $R_\Lambda(\vec p)$ is real and even in $\vec p$. The effective velocity then reads
\begin{subequations}
\begin{align}
    v_n(\vec{x})&=
    a^{-d}\frac{\partial H_\Lambda}{\partial \pi_n(\vec{x})} \nonumber\\
    &=\frac{1}{\rho}\Big[\pi_n(\vec x)+a^d\sum_{\vec y}R_\Lambda(\vec x-\vec y)\,\pi_n(\vec y)\Big],\\
    v_n(\vec p) &=\frac{1}{\rho}\big[1+R_\Lambda(\vec p)\big]\pi_n(\vec p).
\end{align}
\end{subequations}
The concrete lattice form of $R_\Lambda(\vec p)$ is specified in Eq.~\eqref{eq:lattice_regs} below.

We impose transversality of the momentum field $\vec{\pi}$ by demanding 
\begin{align}
    (D_m \pi_m)(\vec{x}) =0,
\end{align}
where $D_m$ denotes the central difference operator
\begin{align}
    (D_m f)(\vec{x}) = \frac{f(\vec{x}+a\hat{e}_m)-f(\vec{x}-a\hat{e}_m)}{2a},
\end{align}
which fulfills, by periodicity, the integration-by-parts (IBP) identity
\begin{align}
    \sum_{\vec{x}} f(\vec{x})\, (D_m g)(\vec{x}) = -\sum_{\vec{x}} g(\vec{x})\, (D_m f)(\vec{x}).
\end{align}
Using this definition, transversality is most conveniently enforced in Fourier space via the transverse projector
\begin{align}
    \mathcal{T}_{mn}(\vec{p}) = \delta_{mn}-\frac{p^c_m p^c_n}{(p^c)^2},
    \qquad p^c_m \equiv \frac{\sin(ap_m)}{a}.
\end{align}
We set $\mathcal{T}_{mn}(\vec p)=\delta_{mn}$ where $\vec p^{\,c}=0$. In particular, $\mathcal{T}_{mn}(\vec 0)=\delta_{mn}$ ensures that the zero mode, i.e., the total momentum $\vec\Pi$, is not removed by the projection. Note that, as in the continuum, the discrete transverse projector is idempotent, self-adjoint, and annihilates lattice gradients, since
\begin{align}
    (D_m \phi)(\vec{p})=i p_m^c \phi(\vec{p})
\end{align}
is parallel to $\vec{p}^{\,c}$. Diagonality of $R_\Lambda$ in Fourier space and transversality of $\vec{\pi}$ imply transversality of $\vec{v}$, i.e., $D_m \pi_m = D_m v_m=0$.

For the dissipative kernel, we replace the Laplacian in Eq.~\eqref{eq:reg_kin_coeff} by the nearest-neighbor Laplacian, with Fourier symbol
\begin{align}
    p^\Laplace_m \equiv \frac{2\sin(a p_m/2)}{a} .
\end{align}
Both regulators are evaluated at this lattice momentum,
\begin{align}
    R_\Lambda(\vec p) = r\big(|\vec p^{\,\Laplace}|/\Lambda\big),
    \qquad
    K_\Lambda(\vec p) = k\big(|\vec p^{\,\Laplace}|/\Lambda\big),
    \label{eq:lattice_regs}
\end{align}
where we employ the regulator~\eqref{eq:reg_family} with $\alpha=1$ for $K_\Lambda$.

\subsection{Discrete equations of motion}
Before discretizing the equations of motion~\eqref{eq:reg_eom}, we write them in the form
\begin{align}
    \partial_t \vec\pi = \partial_t \vec\pi \big|_\text{rev.} + \partial_t \vec\pi \big|_\text{diss./fluct.} \,,
\end{align}
i.e., we split into reversible and combined dissipative-fluctuating contributions, and integrate them alternately using a Lie--Trotter splitting of the spatially discretized stochastic dynamics. Each time step of size $\Delta t$ first advances the dissipative-fluctuating part, which is linear and integrated exactly (see below), and then the reversible part. The transverse projection is applied to the field after the dissipative-fluctuating substep and to the right-hand side of the reversible part in every Runge--Kutta stage. Observables are recorded after the dissipative-fluctuating substep.

\subsubsection{Reversible step}

We first consider the spatial discretization of the reversible part, which is encoded in the Poisson bracket. Deriving the reversible term of the equations of motion~\eqref{eq:eom_general} from the Poisson bracket~\eqref{eq:bracket} requires the product rule, which central differences violate at $O(a^2)$. Discretizing Eq.~\eqref{eq:bracket} directly would thus yield a scheme differing at $O(a^2)$ from the one obtained by discretizing Eq.~\eqref{eq:reg_eom}. We instead perform all steps requiring the product rule in the continuum and discretize the result.
The Poisson bracket of two observables $F[\vec{\pi}]$ and $G[\vec{\pi}]$ is given by
\begin{align}
    \{F,G\} &= \int_{\vec x}\int_{\vec y}
    \frac{\delta F[\vec{\pi}]}{\delta \pi_i(\vec x)}\, \{\pi_i(\vec{x}),\pi_j(\vec{y})\}\,
    \frac{\delta G[\vec{\pi}]}{\delta \pi_j(\vec y)}\\ \nonumber
    &= \int_{\vec x}\frac{\delta F[\vec{\pi}]}{\delta\pi_i(\vec{x})}\frac{\delta G[\vec{\pi}]}{\delta\pi_j(\vec{x})}
      \left(\frac{\partial \pi_j(\vec{x})}{\partial x_i}-\frac{\partial\pi_i(\vec{x})}{\partial x_j}\right)+\dots,
\end{align}
where we have integrated by parts in the second line. The additional terms are proportional to $\nabla\cdot(\delta F/\delta\vec\pi)$ or $\nabla\cdot(\delta G/\delta\vec\pi)$ and thus vanish on the transverse subspace. On the transverse subspace one may thus equally write
\begin{align}
\begin{split}
    &\{\pi_m(\vec{x}),\pi_n(\vec{y})\} \\
    &= \int_{\vec z}\mathcal{T}_{mk}(\vec x-\vec z) \left(\frac{\partial \pi_l(\vec{z})}{\partial z_k} -\frac{\partial \pi_k(\vec{z})}{\partial z_l}\right) \mathcal{T}_{ln}(\vec z-\vec y) \,,
\end{split}
\end{align}
which can be discretized directly as\footnote{Since central differences violate the product rule, the Jacobi identity does not hold for the discrete bracket. We therefore refer to it as an antisymmetric bracket rather than a Poisson bracket. Only antisymmetry is needed for the conservation laws discussed below.}
\begin{align}
    &\{\pi_m(\vec{x}),\pi_n(\vec{y})\}  \label{eq:disc_bracket}\\
    &\to a^d\sum_{\vec z}\mathcal{T}_{mk}(\vec x-\vec z)
      \left(D_k\pi_l-D_l\pi_k\right)(\vec z)\,
      \mathcal{T}_{ln}(\vec z-\vec y) \,.\nonumber
\end{align}
Since $\vec v$ is transverse, the right projector acts trivially, and the reversible step reads
\begin{align}
    \partial_t\pi_m(\vec{x}) \big|_\text{rev.}
    = \mathcal{T}_{ml}\left[ v_n
      \left(D_l\pi_n-D_n\pi_l\right) \right](\vec{x}) \,.
    \label{eq:rev_lattice}
\end{align}

Since the discrete bracket \eqref{eq:disc_bracket} is antisymmetric, the spatially discretized reversible dynamics conserves $H_\Lambda$ exactly, by the same argument as in the continuum. 
Momentum conservation, on the other hand, has to be verified explicitly, since the argument from Sec.~\ref{sec:rev_terms} relies on infinitesimal translational symmetry, which is not available in the lattice-discretized theory. Summing Eq.~\eqref{eq:rev_lattice} over all sites projects onto the zero mode, where $\mathcal{T}_{mn}(\vec 0)=\delta_{mn}$, such that the projector drops out. Hence
\begin{align}
    \partial_t \Pi_m \label{eq:disc_mom_cons_proof}
    &= -a^d \sum_{\vec x} v_n
       \left( D_n \pi_m - D_m \pi_n \right)\\
    &= a^d \sum_{\vec x} \pi_m \, D_n v_n \nonumber\\
    &\quad + \frac{a^{2d}}{\rho} \sum_{\vec x, \vec y}
       \pi_n(\vec x)
       \left[ \frac{\delta_{\vec x \vec y}}{a^d}
       + R_\Lambda(\vec x - \vec y)\right]
       (D_m \pi_n)(\vec{y})\nonumber\\
    &= 0 \,.\nonumber
\end{align}
The first term, obtained by discrete integration by parts, vanishes by transversality of $\vec v$. The second vanishes because $(1+R_\Lambda)D_m$ is antisymmetric, as $R_\Lambda$ is a symmetric convolution that commutes with the antisymmetric $D_m$.

In our numerical implementation, we solve Eq.~\eqref{eq:rev_lattice} using a Runge--Kutta--Fehlberg (4,5) method~\cite{GSL} at fixed step size $\Delta t$. In each Runge--Kutta stage we first evaluate the term in square brackets in position space, and then carry out the transverse projection in Fourier space. Energy conservation is thus violated only by the time-integration error of the Runge--Kutta algorithm. Since the zero mode of the right-hand side vanishes identically for every configuration, $\vec\Pi$ is conserved exactly for any step size.

\subsubsection{Dissipative-fluctuating step}

Let us now consider the dissipative-fluctuating contribution. With the nearest-neighbor Laplacian, the dissipative kernel becomes
\begin{align}
    \Gamma_{mn}(\vec p) = \Gamma(\vec p)\,\delta_{mn},
    \quad
    \Gamma(\vec p) = -\eta\left[1+K_\Lambda(\vec p)\right]
      \big(\vec p^{\,\Laplace}\big)^2 .
\end{align}
As inherited from the continuum, the dissipative kernel is symmetric and negative semi-definite, since $K_\Lambda\geq 0$.

On the transverse subspace, the dissipative-fluctuating system is then given by
\begin{align}
    \partial_t \vec{\pi}(\vec p)
    = \Gamma(\vec p)\,\vec v(\vec p) + \vec\xi(\vec p)
    = -\gamma(\vec p)\,\vec\pi(\vec p) + \vec\xi(\vec p)
\end{align}
with decay rate
\begin{align}
    \gamma(\vec p)
    = \frac{\eta}{\rho}
      \left[1+K_\Lambda(\vec p)\right]
      \left[1+R_\Lambda(\vec p)\right]
      \big(\vec p^{\,\Laplace}\big)^2 .
      \label{eq:decay_rate}
\end{align}
We emphasize that, unlike $p_m^c$, the Fourier symbol $p_m^\Laplace$ does not vanish at the edge of the Brillouin zone, such that every nonzero mode is damped and eventually thermalizes. Due to linearity, the equation can be integrated exactly as
\begin{align}
\begin{split}
    \vec\pi(t_0+\Delta t,\vec p)
    =&\; e^{-\gamma(\vec p)\Delta t}\,\vec\pi(t_0,\vec p)\\
    &+ \sqrt{\frac{\rho T\big(1-e^{-2\gamma(\vec p)\Delta t}\big)}
       {a^d\left(1+R_\Lambda(\vec p)\right)}}
       \;\vec\zeta(\vec p) \,,
\end{split}
\end{align}
where the prefactor of the noise term is fixed by the FDT~\eqref{eq:noise_reg} such that each mode relaxes to its equilibrium variance $\rho T V/[1+R_\Lambda(\vec p)]$, and $\vec\zeta$ denotes normalized transverse noise,
\begin{align}
    \langle\zeta_m(\vec p)\zeta_n^*(\vec p^{\,\prime})\rangle = a^d V\,\mathcal{T}_{mn}(\vec p)\,\delta_{\vec p\vec p^{\,\prime}},
\end{align}
which we generate independently in each step by drawing real Gaussian white noise with unit variance per site and component in position space, followed by Fourier transform. Its longitudinal part is removed by the subsequent transverse projection.
Since both $\gamma(\vec 0)$ and the noise amplitude at $\vec p=0$ vanish, the dissipative-fluctuating step leaves the total momentum $\vec\Pi$ unchanged.

\subsection{Time-step errors}

The Lie--Trotter splitting is first order in $\Delta t$. The dissipative-fluctuating substep leaves $p_\text{eq}$ invariant exactly for any $\Delta t$, and the reversible substep does so in the limit of exact time integration, since it conserves $H_\Lambda$ and phase-space volume (see
Appendix~\ref{sec:fokker_planck}). The composition of the two substeps therefore samples $p_\text{eq}$ up to the error of the Runge--Kutta method, which is of high order in $\Delta t$, making it negligible in practice.

Equal-time observables are unaffected by the splitting up to the Runge--Kutta error. Unequal-time correlation functions (Sec.~\ref{sec:results}), on the other hand, are affected by the splitting error, which modifies the relaxation of modes on which advection and dissipation act on comparable time scales.

Moreover, since the dissipative-fluctuating substep is integrated exactly, it remains stable for any step size, whereas a naive explicit integration would require $\Delta t$ to be smaller than the relaxation time of the fastest mode.

\subsection{Physical parameters, units, and continuum limit}
\label{sec:units}
Setting $k_B=1$, temperature is measured in units of energy and only three independent dimensions of length, mass, and time remain. After reaching convergence in time by tuning $\Delta t\to 0$, the discretized theory has six parameters $\eta, \rho, T, \Lambda, a, L$ and is thus described by $6-3=3$ dimensionless quantities. Since $\Lambda$ sets the physical UV scale of the theory, we measure length in units of $\Lambda^{-1}$, mass in units of $\rho \Lambda^{-d}$, and time in units of the dissipation time $\tau_\Lambda\equiv\rho/(\eta\Lambda^2)$ of a mode at scale $\Lambda$, such that momentum densities are measured in units of $\eta\Lambda$.

In these units, the equations of motion become dimensionless. Denoting dimensionless quantities by a hat, i.e., $\hat{\vec x}=\Lambda\vec x$, $\hat{\vec p}=\vec p/\Lambda$, $\hat t=t/\tau_\Lambda$, and $\hat{\vec\pi}=\vec\pi/(\eta\Lambda)$, the regularized equations of motion~\eqref{eq:reg_eom} read
\begin{align}
    \partial_{\hat t}\hat\pi_l
    = \mathcal{T}_{lm}\bigg[
      &-\hat v_n\left(\frac{\partial\hat\pi_m}{\partial\hat x_n}-\frac{\partial\hat\pi_n}{\partial\hat x_m}\right) \nonumber\\
      &+\hat\nabla^2\big(1+\hat K\big)\hat v_m\bigg] + \hat\xi_l \,,
    \label{eq:eom_dimless}
\end{align}
with $\hat v_m=(1+\hat R)\,\hat\pi_m$, where $\hat R$ and $\hat K$ act in Fourier space as multiplication by $r(|\hat{\vec p}|)$ and $k(|\hat{\vec p}|)$, respectively. Physics enters only in the prefactor of the FDT,
\begin{align}
\begin{split}
    \langle\hat\xi_m(\hat t,\hat{\vec p})\,\hat\xi_n(\hat t',\hat{\vec p}^{\,\prime})\rangle
    =&\; 2\,\mathrm{Re}_\Lambda^2\,\hat{\vec p}^{\,2}\left[1+k(|\hat{\vec p}|)\right]\mathcal{T}_{mn}(\hat{\vec p})\\
    &\times(2\pi)^d\delta(\hat{\vec p}+\hat{\vec p}^{\,\prime})\,\delta(\hat t-\hat t') \,,
\end{split}
\label{eq:noise_dimless}
\end{align}
which contains the square of the thermal Reynolds number at the cutoff scale $\Lambda$~\cite{Bandak_2022},
\begin{equation}
  \mathrm{Re}_\Lambda \equiv \frac{\sqrt{\rho T\Lambda^{d-2}}}{\eta} .
  \label{eq:Re}
\end{equation}
$\mathrm{Re}_\Lambda$ measures the ratio between the dissipative and advective time scales of a mode at scale $\Lambda$, which we illustrate by rewriting it in the typical form $\mathrm{Re}_\Lambda=\frac{\rho \mathsf{v} \ell}{\eta}$ with $\mathsf{v}$ being a typical flow velocity and $\ell$ being a characteristic length scale. Considering a fluid cell of volume $\Lambda^{-d}$, we set the scale $\ell=\Lambda^{-1}$ and obtain the thermal velocity $\mathsf{v}\sim\sqrt{T\Lambda^{d}/\rho}$ from equipartition of energy, yielding Eq.~\eqref{eq:Re}. We emphasize that the quantity $\mathrm{Re}_\Lambda$ characterizes thermal fluctuations at the cutoff scale and should not be confused with the Reynolds number of a macroscopic flow.

In numerical simulations, we have additionally $\Lambda a$ as the UV resolution of the lattice relative to the physical scale, as well as $\Lambda L$ as the system size relative to it. The physical parameters are thus $\mathrm{Re}_\Lambda$ and $\Lambda L$, while $\Lambda a$ controls discretization effects. Convergence to the continuum is reached in the limit $\Lambda a\to 0$ for fixed $\mathrm{Re}_\Lambda$ and $\Lambda L$, hence $N\to\infty$. We adjust $\Delta t$ at every resolution $\Lambda a$ independently until convergence is reached, as described in Sec.~\ref{sec:sim_setup}.

\subsection{Simulation setup}
\label{sec:sim_setup}
We study the renormalization of the shear viscosity $\eta$ for a fluid in thermal equilibrium at temperature $T$ and vanishing total momentum $\vec\Pi=0$. In order to prepare an equilibrium ensemble of the momentum field $\vec{\pi}$, we equilibrate configurations by evolving the system with just the dissipative-fluctuating step, which samples $p_\text{eq}$ exactly, for several relaxation times of the slowest mode. Since the reversible step does not affect the equilibrium distribution, as discussed in Appendix~\ref{sec:fokker_planck}, it can be omitted during equilibration.

Subsequently, we enable the reversible step. We determine $\Delta t$ at fixed physical parameters empirically by reducing it until the correlation functions of Sec.~\ref{sec:results} no longer change within statistical errors. We have in practice found convergence for values of $\Delta t$ that are about a factor of 10 smaller than the (bare) relaxation time of the fastest mode,
\begin{align}
    \Delta t \lesssim 0.1\;
    \frac{\rho}{\eta\,\big[1+k(p^\Laplace_\text{max}/\Lambda)\big]\,(p^\Laplace_\text{max})^2}
\end{align}
with $(p^\Laplace_\text{max})^2=4d/a^2$. This is a conservative choice, since the exact integration removes the stiffness of the dissipative part, as discussed previously, such that the latter imposes no stability constraint on $\Delta t$. %

We evolve the system for a time $t_{\text{sim}}$ that is much larger than the (bare) relaxation time of the slowest mode, and average over independent runs with different random seeds to recover the full thermal ensemble.

\section{Renormalization of the shear viscosity}
\label{sec:results}

Now that we have regularized the theory by introducing a suitable regulator that increases the effective viscosity of UV modes as discussed in Sec.~\ref{sec:reg}, we will study the renormalization of the shear viscosity in $d=2$ and $3$ spatial dimensions. Since the regularized theory has a well-defined continuum limit, we will compare results from direct numerical simulations as discussed in Sec.~\ref{sec:numerics} to (semi-)analytical RG improved one-loop calculations of the renormalized viscosity $\eta_R$ discussed in Sec.~\ref{sec:RGImprov}.

\subsection{Momentum correlator and tree level}
In order to extract the renormalized shear viscosity from numerical simulations, we consider the unequal-time two-point correlation function of the momentum density
\begin{align}
    &C(\Delta t,\vec{p})=\frac{1}{d-1}\sum_{lm} \mathcal{T}_{lm}(\vec{p}) \langle \pi_{l}^{*}(t+\Delta t,\vec{p}) \pi_{m}(t,\vec{p}) \rangle,
\end{align}
where $\langle \cdots \rangle$ denotes the statistical average in the thermal ensemble, and the normalization factor corresponds to an averaging over the $(d-1)$ shear modes. We note that in our numerical simulations, we extract the correlator by averaging over the $2d$ on-axis directions of the wavevector $\vec{p}=\pm p\vec{e}_{m}$, as well as over the evolution time $t$, i.e.
\begin{align}
    &C(\Delta t,p) = \bigg\langle \frac{1}{2d(d-1)} \sum_{\pm} \sum_{l\neq m} %
    \frac{1}{t_{\text{sim}}-\Delta t} \int_0^{t_{\text{sim}}-\Delta t} \hspace{-1.0cm} dt \nonumber \\ 
    &\hspace{2.5cm} \pi_l^*(t+\Delta t,\pm p\vec{e}_m) \pi_l(t,\pm p\vec{e}_m) \bigg\rangle,
\end{align}
to maximize statistics. In practice, the integral is realized as a sum over stored outputs. The remaining average is performed over independent runs with random seeds, from which we also extract statistical error bars as they represent different realizations of the Langevin noise.

Neglecting the non-linear self-advection of the fluid for the moment, the correlation function at tree level is given by
\begin{align}
\label{eq:treelevelCorr}
    C(\Delta t, p) = \rho T V e^{-|\Delta t| / \tau(p)},
\end{align}
where, as in Eq.~\eqref{eq:decay_rate},
\begin{align}
    1/\tau(p)=\gamma(\vec p)=\frac{\eta}{\rho}\left[ 1+K_\Lambda(p) \right]p^2
\end{align}
is the inverse relaxation time of shear modes. We also note for later convenience, that in Fourier space the tree-level correlation function in Eq.~\eqref{eq:treelevelCorr} is given by
\begin{align}
\label{eq:treelevelCorrW}
    C(\omega,p)=\frac{2 \rho T V \tau(p)}{\omega^2\tau(p)^2+1}.
\end{align}

However, due to non-linear interactions, the correlation function receives self-energy corrections which in general exhibit a non-trivial frequency $\omega$ dependence and thus change the analytic structure of the correlation function in frequency space (see, e.g., Ref.~\cite{Kovtun:2012rj}).
These also change the behavior of the correlation function as a function of $\Delta t$. Still, we note that in accordance with the discussion in Sec.~\ref{sec:reg} and \ref{sec:RGImprov} the equal-time value $C(\Delta t=0, p)=\rho T V$ is exact since $H$ is quadratic such that the statics are not renormalized.

\begin{figure}[!tb]
    \centering
    \includegraphics[width=\linewidth]{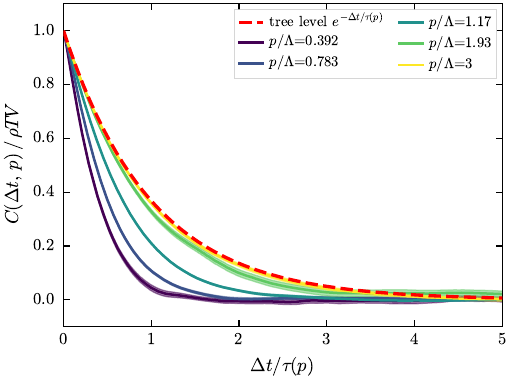}
    \caption{Unequal-time correlation function $C(\Delta t,p)$ of the transverse momentum density for a two-dimensional incompressible fluid extracted from direct numerical simulations with $N=48$, $\mathrm{Re}_\Lambda^{-1} = 0.1$, $\Lambda a = 0.333$. The axes are normalized to the thermal equilibrium value and the tree-level autocorrelation time per $p$, respectively. The red dashed curve shows the tree-level result.}
    \label{fig:2d_correlator}
\end{figure}

We present a typical example of the correlation function in Fig.~\ref{fig:2d_correlator}, where different curves show the correlation function $C(\Delta t,p)/\rho T V$ as a function of $\Delta t/\tau(p)$ for different (lattice) wavenumbers $(p^\Laplace)^2/\Lambda^2 = \frac{4}{\Lambda^2 a^2}\sin^2(\frac{\pi m}{N})$. Shaded bands indicate statistical uncertainties, obtained from the scattering between independent runs as described above. In this context, we use $p$ and $p^\Laplace$ synonymously. While the equal-time value determined by the statics agrees well with the analytical result, one clearly observes how the low-wavenumber modes $p \lesssim \Lambda$ feature a much faster decay compared to the tree-level results shown as the red dashed curve, which signifies the renormalization of the viscosity. However, a careful inspection also reveals deviations from a purely exponential behavior at early times $\Delta t \lesssim \tau(p)$, where the correlation function features additional transient excitations, which can be associated to the non-trivial frequency dependence of the self-energy.%

\subsection{Extraction of the renormalized viscosity}

Since the correlation function $C(\Delta t,p)$ is not just a simple exponential, different extractions of $\tau_R(p)$ that agree with each other at tree level no longer provide the same value, as they exhibit different sensitivity to the behavior of the correlation function in the complex frequency plane. Here we explore three different extraction procedures, namely 1) the integrated autocorrelation time and 2) fits in Fourier space, both of which aim to characterize the low-frequency behavior of $C(\omega,p)$, as well as 3) the exponential autocorrelation time, which aims to extract the imaginary part $\mathrm{Im}(\omega)$ of the closest singularity of $C(\omega,p)$ in the complex frequency plane. Below, we briefly explain the three methods using the example of the correlation function in Fig.~\ref{fig:2d_correlator}.

For such a renormalized relaxation time $\tau_R(p)$ extracted from $C(\Delta t,p)$, we define the renormalized shear viscosity as
\begin{align}
\label{eq:EtaRFromTauRDef}
    \eta_{R}(p)=\frac{\rho}{\tau_R(p)p^2(1+K_{\Lambda}(p))},
\end{align}
which reduces to the bare viscosity $\eta$ at tree level, where $\tau_R(p)=\tau(p)$. We denote the relaxation times obtained from the three procedures by $\tau_R^\text{int}$, $\tau_R^\text{FT}$, and $\tau_R^\text{exp}$, and the corresponding viscosities by $\eta_R^\text{int}$, $\eta_R^\text{FT}$, and $\eta_R^\text{exp}$. 

\subsubsection{Integrated autocorrelation time}

\begin{figure}[!tb]
    \centering
    \includegraphics[width=\linewidth]{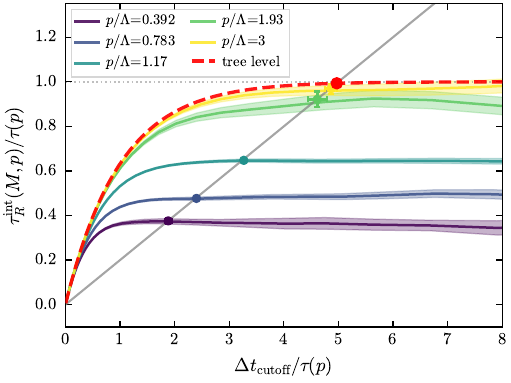}
    \caption{Running estimate $\tau_R^\text{int}(M,p)/\tau(p)$ of the integrated autocorrelation time as a function of the window size $\Delta t_\text{cutoff}/\tau(p)$ for the correlation functions in Fig.~\ref{fig:2d_correlator}, with shaded bands indicating statistical uncertainties and the red dashed curve showing the tree-level result. The extracted values $\tau_R^\text{int}(p)$ (points with error bars) are given by the intersections with the grey line of slope $1/c$ ($c=5$), where the Madras--Sokal criterion~\eqref{eq:sokal_criterion} is first fulfilled.}
    \label{fig:sokal}
\end{figure}

By considering the integrated autocorrelation time 
\begin{align}
    \tau_{R}^\text{int}(p)=\int_{0}^{\infty}\, d\Delta t~ \frac{C(\Delta t,p)}{C(\Delta t=0,p)},
\end{align}
where $C(\Delta t=0,p)=\rho T V$  is the equal-time correlator, we can access the zero-frequency limit
\begin{align}
    \tau_R^\text{int}(p)=\frac{C(\omega=0,p)}{2C(\Delta t=0,p)}
    \label{eq:tauint_omega0}
\end{align}
of the Fourier space correlation function. Since, in practice, the numerical integration of the correlation function suffers from large statistical errors at late times $\Delta t \gtrsim \tau_{R}^\text{int}(p)$, we employ the  self-consistent Madras--Sokal windowing procedure~\cite{madrasPivotAlgorithmHighly1988} (see also Ref.~\cite{sokalMonteCarloMethods1997} for a pedagogical introduction). We approximate the integral by a sum over the stored outputs, spaced by the output interval $\delta t$, and truncate it after $M$ data points,
\begin{align}
    \tau_R^\text{int}(M,p) \approx \frac{\delta t}{2} + \delta t \sum_{n=1}^M \frac{C(n\,\delta t, p)}{C(\Delta t=0, p)},
\end{align}
where $M$ has to be chosen large enough that a plateau is reached but small enough to prevent drifting due to statistical noise of the correlation function. Since the integral converges after a few relaxation times, we self-consistently choose $M$, following Refs.~\cite{madrasPivotAlgorithmHighly1988, sokalMonteCarloMethods1997}, as the smallest integer such that
\begin{align}
    \Delta t_{\rm cutoff}=M\delta t \geq c \tau_R^\text{int}(M).
    \label{eq:sokal_criterion}
\end{align}
where the constant $c$ has to be sufficiently large to reach convergence of the integral. In practice, we choose $c=5$. We show an example of this extraction method in Fig.~\ref{fig:sokal}, where the different curves show the running estimate $\tau_R^\text{int}(M)/\tau(p)$ as a function of the cutoff time $\Delta t_{\rm cutoff}(p)/\tau(p)$ for the different wavenumbers shown in Fig.~\ref{fig:2d_correlator}. Data points mark the self-consistently determined values of $\tau_R^\text{int}(p)$. We estimate statistical error bars by performing the extraction of $\tau_R^\text{int}$ for each run, with the final result given as the mean and error of the mean value over runs. The choice of $c$ introduces a systematic uncertainty, since slow tails on top of a rapid decay are underestimated for too small windows. For the value of $c$ used here, the running estimates in Fig.~\ref{fig:sokal} have reached a plateau at the cutoff, and extending the window further changes them by less than their statistical uncertainty.

\begin{figure}[t]
    \centering
    \includegraphics[width=\linewidth]{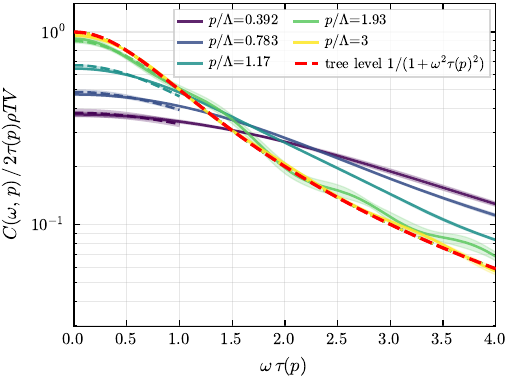}
    \caption{Extraction of the relaxation time $\tau_R^\text{FT}(p)$ from a fit of the Fourier-space correlation function $C(\omega,p)/[2\tau(p)\rho TV]$ as a function of $\omega\tau(p)$. The red dashed curve shows the tree-level result, Eq.~\eqref{eq:treelevelCorrW}. Dashed lines with shaded bands show fits to a Lorentzian as described in the main text. 
    }
    \label{fig:fourier_fit}
\end{figure}

\subsubsection{Fourier fit}
By performing a discrete cosine transform of the (real, even) unequal-time correlation function, we obtain the correlation function in Fourier space as
\begin{align}
    C(\omega,p)=\int_{0}^{\infty}\,d\Delta t\,2\cos(\omega \Delta t)\,C(\Delta t, p)
\end{align}
where we again evaluate the integral as a sum over outputs spaced by $\delta t$. 
By investigating the low frequency behavior, we then extract the renormalized relaxation time $\tau_R^\text{FT}(p)$, by performing a Lorentzian fit as in Eq.~\eqref{eq:treelevelCorrW}, treating
the width $\tau_R^\text{FT}(p)$ as a free parameter. We Fourier transform the correlator for each run independently and then average the result, which yields a statistical error at each frequency.  Systematic errors are estimated by varying the upper bound of the fit window  and give the total error as the quadrature sum of systematic and statistical errors.

We show the results for the Fourier-transformed correlation function in Fig.~\ref{fig:fourier_fit}, where again different curves correspond to the different wavenumbers in Fig.~\ref{fig:2d_correlator}. While for $p\gg \Lambda$ the correlator agrees with the tree-level result in Eq.~\eqref{eq:treelevelCorrW}, the renormalization of the viscosity for low-wavenumber modes $p \lesssim \Lambda$ results in a sizeable suppression of the low-frequency peak. From the fitted width, we obtain $\eta_R^\text{FT}(p)$ via Eq.~\eqref{eq:EtaRFromTauRDef}.

\begin{figure}[t]
    \centering
    \includegraphics[width=\linewidth]{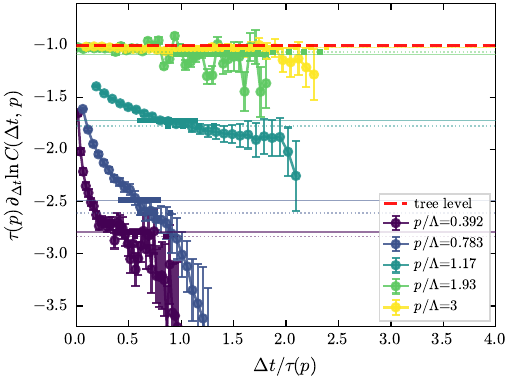}
    \caption{Logarithmic derivative $\tau(p)\,\partial_{\Delta t}\ln C(\Delta t,p)$ as a function of $\Delta t/\tau(p)$ for the correlation functions in Fig.~\ref{fig:2d_correlator}. At tree level, it equals $-1$ (red dashed line). Solid (dotted) horizontal lines show the fit values $-\tau(p)/\tau_R^\text{exp}(p)=-\eta_R^\text{exp}(p)/\eta$ obtained in the self-consistently determined windows $[\tau_R^\text{exp},2\tau_R^\text{exp}]$ ($[1.5\tau_R^\text{exp},2.5\tau_R^\text{exp}]$). The steep drop at late times signals the first zero crossing of $C(\Delta t,p)$ due to limited statistics.}
    \label{fig:logderiv}
\end{figure}

\subsubsection{Exponential auto-correlation time}
With the exponential auto-correlation time $\tau_R^\text{exp}(p)$ one aims to extract the late time exponential decay of the correlation function
\begin{align}
    C(\Delta t,p) \propto e^{-|\Delta t|/\tau_R^\text{exp}(p)}, \label{eq:exp_decay}
\end{align}
Evidently, the relaxation time $\tau_R^\text{exp}(p)$ can be extracted independently of the overall amplitude by considering the logarithmic derivative
\begin{align}
    \frac{\partial \ln C(\Delta t,p)}{\partial \Delta t} = -\frac{1}{\tau_R^\text{exp}(p)} \,, \quad \Delta t \to \infty \,. \label{eq:C_log_deriv}
\end{align}
We compute the logarithmic derivative by finite differences between outputs separated by $3\delta t$ and fit a constant $-1/\tau_R^\text{exp}$ to Eq.~\eqref{eq:C_log_deriv} for each run, as illustrated in Fig.~\ref{fig:logderiv}. We see that the data shows a plateau for intermediate times, with initial transients dominating before and noise dominating after the plateau.

Inspired by the Madras--Sokal windowing procedure, we determine the fit interval $[\tau_R^\text{exp},2\tau_R^\text{exp}]$ by requiring self-consistency with the fit value $-1/\tau_R^\text{exp}$. We estimate the statistical error by the run-to-run scattering of the resulting fit values. To estimate the systematic error, we repeat the procedure, but require self-consistency of the shifted fit interval $[1.5\tau_R^\text{exp},2.5\tau_R^\text{exp}]$. The fit values obtained in the two self-consistently determined windows are shown as solid and dotted horizontal lines, respectively. Our estimate for the systematic error given by the difference of the values obtained from both fit windows is comparable to the statistical error resulting from the run-to-run scattering of the fits, leading to a rather precise extraction of the renormalized viscosity.

Clearly, a limitation of this method is the appearance of 
zero-crossings of $C(\Delta t, p)$ due to an exponentially small signal and insufficient statistics at large time differences $\Delta t$. 
Since $C(\Delta t, p)$ itself is strictly positive before the first zero-crossing, the slope at the zero-crossing must be negative, which means that the logarithmic derivative~\eqref{eq:C_log_deriv} approaches $-\infty$, as visible in Fig.~\ref{fig:logderiv}. Since the signal decays exponentially, avoiding such zero-crossings requires statistics that grow exponentially in $\Delta t$.  This method is therefore restricted to time differences $\Delta t$ below the first zero-crossing of $C(\Delta t, p)$. More specifically, the upper bound of our second fit interval ($2.5\tau_R$) must always be smaller than the $\Delta t$ of the first zero-crossing of the correlation function.

\subsection{Two dimensions}
\label{sec:results_2d}

\begin{figure}[tb]
    \centering
    \includegraphics[width=0.9\linewidth]{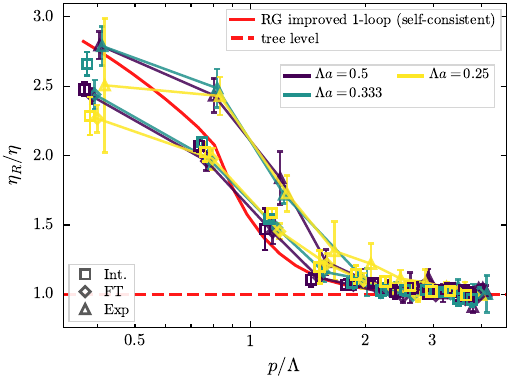}
    \caption{Renormalized shear viscosity $\eta_R(p)/\eta$ of a two-dimensional fluid at $\mathrm{Re}_\Lambda^{-1}=0.1$ and fixed $\Lambda L=16$, for different lattice spacings. Symbols denote the extraction method. The solid red line shows the self-consistent RG improved result of Sec.~\ref{sec:self_cons_RG_improv}, and the dashed red line the bare viscosity.}
    \label{fig:2d_cont_limit}
\end{figure}

We will first investigate the behavior of two-dimensional fluids and explicitly verify that the regularized theory admits a well defined continuum limit.  To do so, we take the lattice spacing  $\Lambda a\to 0$ for a fixed system size $\Lambda L=16$, as demonstrated exemplarily in Fig.~\ref{fig:2d_cont_limit}, where we present results for the renormalized shear viscosity $\eta_R$ as a function of wavenumber $p/\Lambda$ for an inverse Reynolds number $\mathrm{Re}_\Lambda^{-1}=0.1$. Different colored curves in Fig.~\ref{fig:2d_cont_limit} correspond to the results for different lattice spacings, while different symbols correspond to different extraction methods, i.e. $\tau_{R}^{\rm int}$ (boxes), $\tau_{R}^\text{FT}$ (diamonds) and $\tau_{R}^{\rm exp}$ (triangles). 

Generally, we find that the integrated autocorrelation time and Fourier analysis provide similar estimates for the renormalized viscosity, while the logarithmic derivative yields systematically larger values for $p\lesssim\Lambda$. While for $p \lesssim \Lambda$ the renormalized shear viscosity experiences an approximately logarithmic rise for small wave-numbers, we find that, for large wavenumbers $p \gtrsim \Lambda$, it converges towards the bare shear viscosity indicated by the red dashed line. Most importantly, the data agrees well for different lattice spacings $\Lambda a$, indicating that convergence to the continuum has been reached. 

We also compare our results from direct numerical simulations to continuum calculations using the self-consistent RG improved calculation (see Sec.~\ref{sec:self_cons_RG_improv}), which are indicated in Fig.~\ref{fig:2d_cont_limit} as a solid red line. Even though the RG improved perturbative calculation slightly overestimates the renormalization of the viscosity at small wavenumbers, we find good agreement between the lattice simulations and our (semi-) analytic calculations in the continuum.

\begin{figure}[t]
    \centering
    \includegraphics[width=0.9\linewidth]{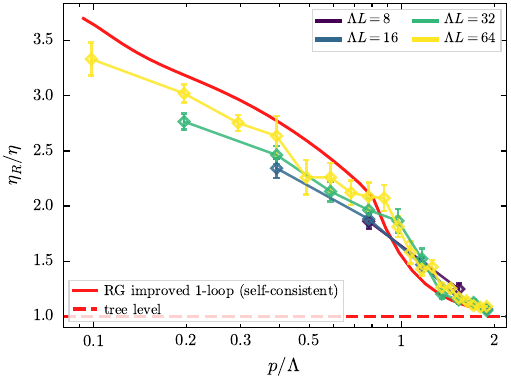}
    \caption{System-size dependence of the renormalized shear viscosity $\eta_R(p)/\eta$ of a two-dimensional fluid at fixed $\Lambda a=0.5$ and $\mathrm{Re}_\Lambda^{-1}=0.1$, extracted from the Fourier fit. The solid red line shows the self-consistent RG improved calculation as described in Sec.~\ref{sec:self_cons_RG_improv}.}
    \label{fig:2d_system_size}
\end{figure}

The renormalized shear viscosity in a more narrow sense is defined in the limit $p\to 0$, where, in two spatial dimensions, RG improved perturbative calculations suggest a logarithmic system size dependence of the form (cf.~Eq.~\eqref{eq:etaR_result_2d})
\begin{align}
\label{eq:2DetaR_Re}
    \frac{\eta_{R}}{\eta} = \sqrt{1 + \frac{\mathrm{Re}_\Lambda^2}{8\pi} \ln (\sqrt{e}\Lambda L/2\pi)},
\end{align}
where with our regularization, the relevant UV cutoff is set by the regularization scale $\Lambda$, while the logarithmic dependence on the system size $L$ is expected to persist, as the smallest wavenumber $p_{\rm min}$ of the lattice discretized theory provides a natural IR cutoff $\mu=p_{\rm min}=2\pi/L$.\footnote{Note that in Eq.~\eqref{eq:2DetaR_Re} we have set $\alpha=1$ as this is the case for the regulator employed in our numerical simulations. However, since the hard IR cutoff $\mu$ employed in the continuum calculation is not exactly equivalent to the finite lattice volume $L$ of our simulation, one should anticipate possible variations of the scale of the logarithm.} Behavior consistent with Eq.~\eqref{eq:2DetaR_Re} was also recently reported in~\cite{Nakano_2026}, where the UV scale of the logarithm was treated as a fit parameter set by the lattice discretization.

While for $\mathrm{Re}_\Lambda^2 \ln (\sqrt{e}\Lambda L/2\pi) \ll 1$, the renormalized viscosity in Eq.~\eqref{eq:2DetaR_Re} shows a logarithmic system size dependence
\begin{align}
    \eta_{R} \approx \eta + \frac{\rho T}{16\pi \eta}\ln (\sqrt{e}\Lambda L/2\pi)
\end{align}
as is well established in the literature~\cite{Forster:1977zz}, Eq.~\eqref{eq:2DetaR_Re} also predicts a change in behavior for $\mathrm{Re}_\Lambda^2 \ln (\sqrt{e}\Lambda L/2\pi) \gg 1$, where
\begin{align}
\label{eq:etaR2D_98}
 \eta_R \approx \sqrt{\frac{\rho T}{8\pi}\ln (\sqrt{e}\Lambda L/2\pi)}
\end{align}
shows a weaker logarithmic divergence with increasing system size and becomes independent of the bare viscosity $\eta$.

\begin{figure}
    \centering
    \includegraphics[width=\linewidth]{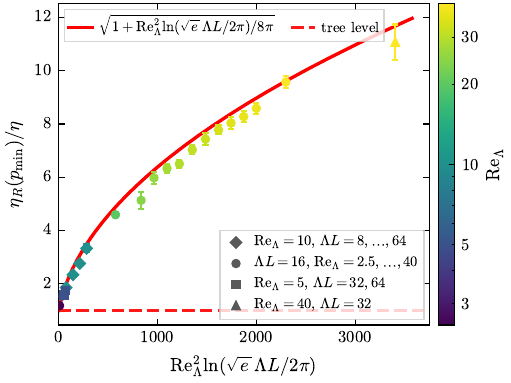}
    \caption{Renormalized shear viscosity of the lowest mode, $\eta_R(p_\text{min})/\eta$ with $p_\text{min}=2\pi/L$, of a two-dimensional fluid as a function of $\mathrm{Re}_\Lambda^2\ln(\sqrt{e}\Lambda L/2\pi)$, extracted from  the FT method at $\Lambda a=0.5$. Symbols denote series of simulations at fixed $\mathrm{Re}_\Lambda$ and varying $\Lambda L$ or vice versa, as indicated in the legend, and colors indicate $\mathrm{Re}_\Lambda$. The solid red line shows the RG improved prediction, Eq.~\eqref{eq:2DetaR_Re}, and the dashed red line the bare viscosity.}
    \label{fig:2d_Visc_Divergence}
\end{figure}

We investigate the system size dependence of the renormalized viscosity in Fig.~\ref{fig:2d_system_size}, where we study the wavenumber dependence of the renormalized viscosity  $\eta_{R}(p)$ for different system sizes $\Lambda L=8 \cdots 64$ for fixed lattice spacing $\Lambda a=0.5$ close to the continuum limit. While the results for different system size approximately agree with each other for larger wavenumbers, larger systems provide access to smaller wavenumbers   $p \ll \Lambda$, for which the renormalized viscosity $\eta_{R}(p)$ experiences a logarithmic rise. When considering the smallest wave-number $p_{\rm min}=2\pi/L$ accessible in a system of size $L$, this results in a logarithmic increase of the viscosity for increasing system size, as can be seen from Fig.~\ref{fig:2d_Visc_Divergence}, where we show the results for $\eta_{R}(p_{\rm min})/\eta$ as a function of $\mathrm{Re}_\Lambda^2\ln(\sqrt{e}\Lambda L/2\pi)$.

Strikingly, as suggested by Eq.~\eqref{eq:2DetaR_Re},  the data in Fig.~\ref{fig:2d_Visc_Divergence} collapses onto a single curve when plotted against the scaling variable $\mathrm{Re}_\Lambda^2\ln(\sqrt{e}\Lambda L/2\pi)$, irrespective of varying the system size at fixed Reynolds number $\mathrm{Re}_\Lambda$ or the Reynolds number at fixed system size $\Lambda L$. 
By considering variations of $\Re_{\Lambda}$ and $\Lambda L$, the data in Fig.~\ref{fig:2d_Visc_Divergence} extends from the perturbative regime $\mathrm{Re}_\Lambda^2\ln(\sqrt{e}\Lambda L/2\pi)\lesssim 1$, where $\eta_{R}/\eta \approx 1$ far into the regime $\mathrm{Re}_\Lambda^2\ln(\sqrt{e}\Lambda L/2\pi)\gg 1$, where the renormalized viscosity becomes independent of the bare viscosity and grows only as the square root of the logarithm of the system size (cf.~\eqref{eq:etaR2D_98}). The RG improved prediction, shown as a solid red line, describes this curve without any free parameter, with less than 10\%  deviations. 

\subsection{Three dimensions}

\begin{figure}[t]
    \centering
    \includegraphics[width=\linewidth]{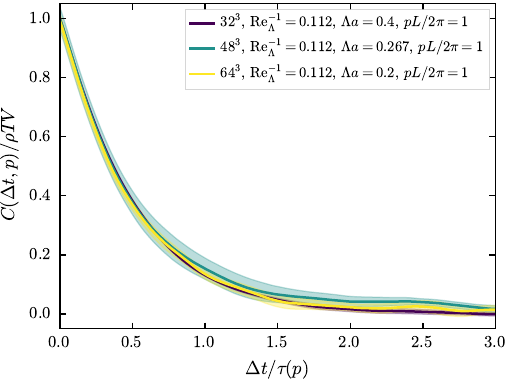}
    \caption{Unequal-time correlation function $C(\Delta t,p)/\rho TV$ of the lowest nonzero mode, $pL/2\pi=1$ of a three-dimensional fluid as function of $\Delta t/\tau(p)$, for different lattice spacings at fixed $\Lambda L=12.8$ and $\mathrm{Re}_\Lambda^{-1}=0.112$. Shaded bands indicate statistical uncertainties. The results agree within errors, indicating convergence to the continuum limit.}
    \label{fig:3d_cont_limit}
\end{figure}

Next we will consider incompressible fluids in three spatial dimensions, starting with numerical results for the autocorrelation function $C(\Delta t, p)$ which are presented in Fig.~\ref{fig:3d_cont_limit}, for  different lattice discretizations $\Lambda a$. Since the correlation functions for different $\Lambda a$ agree well within error bars, we conclude that convergence towards the continuum has been reached for the considered values of $\Lambda a$. 

Generally, the correlation function in Fig.~\ref{fig:3d_cont_limit} shows the same qualitative behavior as in two spatial dimensions. We thus proceed in analogy to the two-dimensional case, and extract the renormalized shear viscosity as a function of wavenumber $p/\Lambda$  from the integrated autocorrelation time, the analysis in Fourier space and the exponential autocorrelation time. 

\begin{figure*}[t]
    \centering
    \begin{minipage}{.49\textwidth}
        \centering
        {(a) $\mathrm{Re}_\Lambda^{-1} = 0.79$}
        \includegraphics[width=\linewidth]{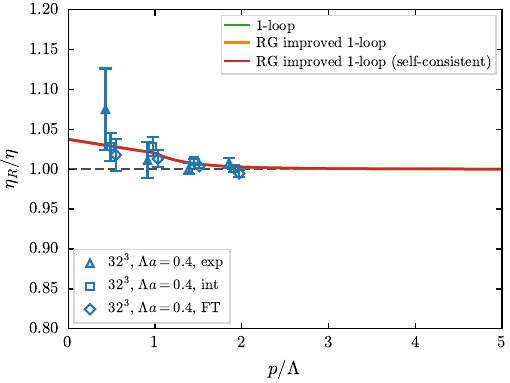}
        {(c) $\mathrm{Re}_\Lambda^{-1} = 0.063$}
        \includegraphics[width=\linewidth]{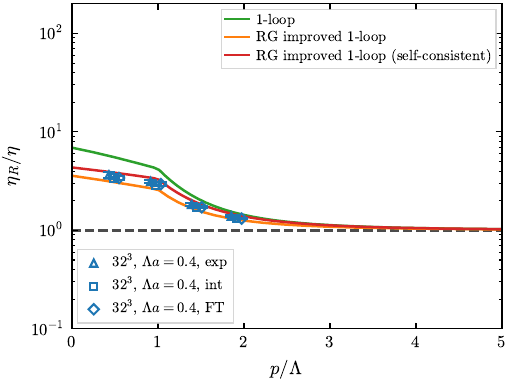}
    \end{minipage}
    \begin{minipage}{.49\textwidth}
        \centering
        {(b) $\mathrm{Re}_\Lambda^{-1} = 0.11$}
        \includegraphics[width=\linewidth]{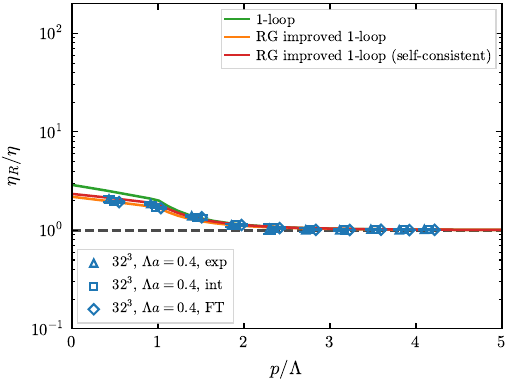}
        {(d) $\mathrm{Re}_\Lambda^{-1} = 0.011$}
        \includegraphics[width=\linewidth]{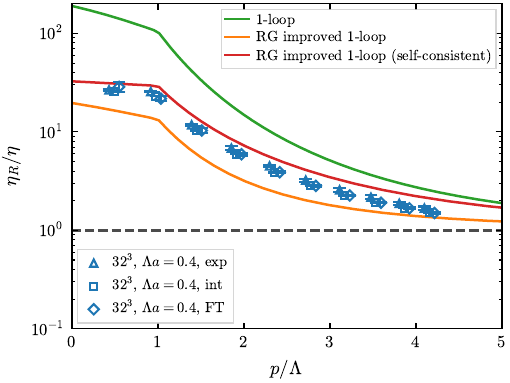}
    \end{minipage}
    \caption{Wavenumber-dependent renormalized viscosity $\eta_R(p)/\eta$ of a three-dimensional fluid at $\Lambda a=0.4$ and $\Lambda L=12.8$, for small (a) to large (d) thermal Reynolds numbers. Blue symbols show direct numerical simulations for the three extraction methods.  Results for the extracted shear viscosity from the `exp' (`FT') method are offset slightly to the left (to the right) to improve readability. Solid lines show one-loop perturbation theory (green) and the RG improved calculation without (orange) and with (red) self-consistent wavenumber dependence. The grey dashed line indicates the bare viscosity.}
    \label{fig:3d_k_dep}
\end{figure*}

We present the results in Fig.~\ref{fig:3d_k_dep},  where different panels show the result for different Reynolds numbers $\text{Re}_{\Lambda}$. Since in three dimensions, the renormalized shear viscosity is expected to remain finite in the infinite volume limit, we can directly compare the numerical simulation results with our (semi-)analytical calculations in the infinite volume limit.\footnote{We have also investigated the volume dependence explicitly by introducing a hard IR cutoff $\mu = 2\pi/L$, as a proxy for the finite volume $(Na)^3$ of our numerical simulations. Such finite-volume corrections only cause a small systematic reduction of $\eta_R$ by about $10\%$ for the smallest viscosities.} Different curves in Fig.~\ref{fig:3d_k_dep} correspond to the results obtained from direct numerical simulations (blue points), one-loop perturbation theory (green), RG improved perturbation theory (orange) and RG improved perturbation theory with a self-consistent wave-number dependence of the viscosity (red). Different symbols correspond again to different extraction methods for the renormalized viscosities, where in contrast to the two dimensional case the different methods yield consistent results within errors. 

Generally, all curves show a wavenumber dependent renormalization of the viscosity $\eta_{R}(p)$, which increases for smaller inverse Reynolds numbers $\text{Re}_{\Lambda}^{-1}$ going from panel (a) to (d).  While for $p \gg \Lambda$ the viscosity $\eta_{R}(p)\simeq \eta$ is not significantly renormalized, it experiences a rapid rise for $p \lesssim \Lambda$, where the (semi-)analytical curves feature a cusp reflecting the kink of the regulator~\eqref{eq:reg_family} at $p/\Lambda=1$, where its derivative is discontinuous. 

By using the Reynolds number~\eqref{eq:Re}, the RG improved perturbative result in~\eqref{eq:etaR_result} can be cast into the following dimensionless form
\begin{align}
\label{eq:RGImpEtaRRe}
    \frac{\eta_R}{\eta} = \sqrt{1+2c_d \mathrm{Re}_\Lambda^2},
\end{align}
where for our regulator~\eqref{eq:reg_family} with $\alpha=1$ we obtain the coefficient $c_3=7/(30 \pi^2)\approx 0.0236$ in three dimensions. By comparing the magnitude of the bare viscosity with the perturbative correction, we can thus  estimate that perturbation theory should be accurate for $\mathrm{Re}_\Lambda^{-1} \gg \sqrt{2c_d} \approx 0.217$. While this criterion is well satisfied in Fig.~\ref{fig:3d_k_dep}(a), where the one-loop result is in good agreement with the self-consistent RG improved calculation, one also observes how one-loop perturbation theory over-estimates the renormalization of the shear viscosity for smaller inverse Reynolds numbers as in Figs.~\ref{fig:3d_k_dep}(b)--(d), while the self-consistent RG improved calculation still provides a satisfactory description of the numerical simulations.

\begin{figure}[t]
    \centering
    \includegraphics[width=\linewidth]{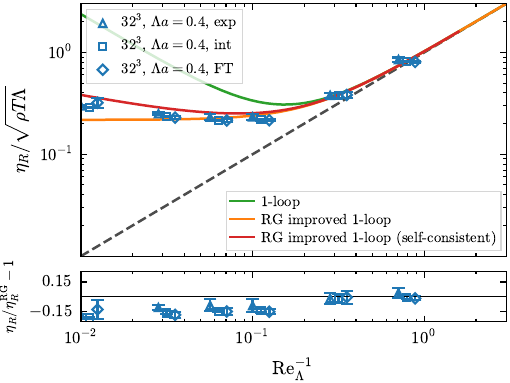}
    \caption{Dimensionless renormalized viscosity $\eta_R/\sqrt{\rho T\Lambda}=\mathrm{Re}_R^{-1}$ of the lowest nonzero mode, $pL/2\pi=1$, as a function of $\mathrm{Re}_\Lambda^{-1}=\eta/\sqrt{\rho T\Lambda}$ in three dimensions, at $\Lambda a=0.4$ and $\Lambda L=12.8$. Symbols and solid lines are as in Fig.~\ref{fig:3d_k_dep}, while the black dashed line indicates $\eta_R=\eta$. The lower panel shows the relative deviation $\eta_R/\eta_R^\text{RG}-1$ of the simulation results from the self-consistent RG improved calculation.}
    \label{fig:eta-ratio-vs-inv-re}
\end{figure}

Since in three spatial dimensions, the renormalization of the viscosity does not suffer from infrared divergencies, we expect $\eta_{R}(p)$ to be finite in the limit $p \to 0$, which allows us to define a renormalized viscosity as $\eta_{R}=\lim_{p \to 0}\eta_{R}(p)$. In order to further analyze the dependence on the inverse Reynolds number $\mathrm{Re}_\Lambda^{-1}$, we infer the renormalized $\mathrm{Re}_R^{-1}=\eta_R/\sqrt{\rho T\Lambda}$ from the viscosity $\eta_{R}(p_{\rm min})$ of the lowest wavenumber $(p_{\rm min}=2\pi/L)$ mode, and investigate it as a function of the bare inverse Reynolds number $\mathrm{Re}_\Lambda^{-1}$, as depicted in Fig.~\ref{fig:eta-ratio-vs-inv-re}.
While for large inverse Reynolds number $\mathrm{Re}_\Lambda^{-1}\gtrsim 1$, the renormalization of the shear viscosity is small such that $\eta_{R} \approx \eta$, our numerical simulations show a minimum of the renormalized viscosity $\eta_{R}^{\rm min}\approx 0.25\sqrt{\rho T \Lambda}$ for bare values  $\mathrm{Re}_\Lambda^{-1}\sim 0.1$, consistent with the lower bound on the shear viscosity discussed in Refs.~\cite{Kovtun:2011np,Chafin:2012eq}. While one loop perturbation theory (green) clearly over-estimates the renormalization of the shear viscosity for small inverse Reynolds number $\text{Re}_{\Lambda}^{-1} \lesssim 0.5$, 
the RG improved calculation in Eq.~(\ref{eq:RGImpEtaRRe}) (orange) remains accurate up to the minimum of the renormalized viscosity. Self-consistent RG improved calculations (red) are even able to capture the increase of the renormalized viscosity for very small (bare) viscosities with deviations of at most $15\%$, as can be seen from the lower panel of Fig.~\ref{fig:eta-ratio-vs-inv-re}, which shows the relative deviation of the simulation results from the self-consistent RG improved calculation.

While in Ref.~\cite{Kovtun:2011np} the minimum of the renormalized viscosity $\eta_R \gtrsim 0.25 \sqrt{\rho T \Lambda}$ has been  interpreted as an absolute lower limit on viscosity caused by thermal fluctuations, we note that one can also give this result a complementary interpretation regarding the range of applicability of hydrodynamics as an effective field theory. Since in a real world experiment, one measures the renormalized viscosity $\eta_R$, and not $\eta$, this means that the bare coupling $\eta$ of the effective theory of stochastic hydrodynamics should be tuned so that $\eta_R(\eta,\Lambda)$ matches the experimental value. The lower bound on $\eta_R$ for fixed $\Lambda$ from Fig.~\ref{fig:eta-ratio-vs-inv-re} is thus equivalent to an upper bound on $\Lambda$ for fixed $\eta_R$,\footnote{Here we have used the result from Fig.~\ref{fig:eta-ratio-vs-inv-re} that $1/\mathrm{Re}_R^\text{max} \approx 0.25$ for the regulator \eqref{eq:reg_family} with $\alpha=1$ in $d=3$ dimensions.}
\begin{equation}
    \Lambda \lesssim \frac{16 \eta_R^2}{\rho T} \equiv \Lambda_\text{max}. \label{eq:lamBound}
\end{equation}
Stated differently, for any physical system with a given $\eta_{R}$, the stochastic Navier--Stokes equation \eqref{eq:incompr_eom} stops being a valid description beyond $\Lambda_\text{max}$, since for larger UV cut-offs no choice of the bare viscosity $\eta$ can reproduce the measured viscosity $\eta_R$.

\section{Epilogue}
\label{sec:concl_outlook}

\begin{figure*}
    \centering
    \includegraphics[width=0.75\linewidth]{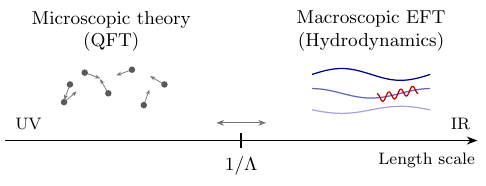}
    \caption{ Stochastic hydrodynamics as an effective field theory, defined with a cutoff $\Lambda$ that separates the microscopic and macroscopic scales. Within stochastic hydrodynamics fluctuations can be assumed to be integrated out down to the scale $\Lambda$, i.e., $1/\Lambda$ can be interpreted as a coarse-graining length. A description in terms of (fluctuating) hydrodynamics is expected to apply for $1/\Lambda \gg l_{mic}$, where $l_{mic}$ denotes some characteristic microscopic scale such as the mean-free path in a dilute gas or $\hbar c/\pi T$ (with $k_B=1$) in a strongly coupled QFT at finite temperature.}
    \label{fig:scales}
\end{figure*}
Stochastic hydrodynamics provides a universal low-energy effective theory for the dynamics of interacting many-body systems on large time and long distance scales. Naturally this effective theory is only properly defined in conjunction with an ultra-violet regularization, which suppresses thermal fluctuations of the hydrodynamic fields on short distance scales $\lesssim 1/\Lambda$. Since the theory is non-linear, the choice of regularization and cut-off scale does affect observable macroscopic properties, such as for example the renormalized shear viscosity $\eta_R(\eta_\text{bare},\Lambda)$ of the fluid. However, as illustrated in Fig.~\ref{fig:scales}, the dynamics below the short distance cut-off scale should be described by the underlying microscopic theory, e.g. a quantum field theory (QFT). Supplemented with a suitable infrared regulator that suppresses long distance hydrodynamic fluctuations in the underlying microscopic theory, the microscopic description then provides the bare parameters of the effective field theory, e.g. $\eta_\text{bare}(\Lambda)$, in such a way that physical observables, such as $\eta_R(\eta_\text{bare}(\Lambda),\Lambda)$ become independent of the artificial cut-off scale $\Lambda$ separating between the microscopic and macroscopic description. 

Evidently, for such a description to hold, the cut-off scale $1/\Lambda$ should be sufficiently large for hydrodynamics to be applicable, i.e. for example for a weakly interacting system $1/\Lambda$ should be much larger than the mean-free path $\lambda_{\rm mfp}$. Below this scale a consistent microscopic description including thermal fluctuations becomes inevitable. While the development of a consistent EFT perspective, matching the IR cut-off of the microscopic theory with the UV cut-off of the EFT, remains an outstanding task, in this work
we proposed UV regularizations of the Navier--Stokes equation with thermal noise that preserve all symmetries. More specifically, we have considered (i) introducing a mass-like UV regulator $R_{\Lambda}$ to the Hamiltonian and (ii) introducing a UV regulator $K_{\Lambda}$ to the viscosity, which both diverge for large wavenumbers $p\to\infty$. The stochastic noise is then fixed by the FDT, and the Poisson bracket is left untouched. By construction, this approach preserves energy and momentum conservation, Galilean invariance, Onsager reciprocity and the stationarity of the equilibrium distribution.

We have shown explicitly that the viscosity regulator $K_\Lambda$ alone renders the one-loop correction to the shear viscosity UV finite. In the sharp-cutoff limit it reproduces the classic result of Forster, Nelson and Stephen~\cite{Forster:1977zz}. A static regulator $R_\Lambda$, by contrast, cannot realize a sharp cutoff. Inspired by the real-time functional renormalization group (FRG) approach developed in Refs.~\cite{Roth:2024rbi,Roth:2024hcu}, we derived an RG improved expression for the renormalized viscosity, Eq.~\eqref{eq:etaR_result}, together with a self-consistent extension for its full wavenumber dependence.

Because the physical cutoff $\Lambda$ is independent of the lattice spacing $a$, the regularized theory admits a genuine continuum limit $\Lambda a \to 0$ at fixed thermal Reynolds number $\mathrm{Re}_\Lambda$ and system size $\Lambda L$. Our structure-preserving discretization conserves momentum exactly and energy up to the Runge--Kutta error, while the dissipative-fluctuating step samples the equilibrium distribution exactly.

We explicitly verified the convergence to the continuum limit, and compared our results from direct numerical simulations to RG improved perturbative calculations.  In two dimensions, we observed the expected logarithmic rise of $\eta_R(p)$ at small wavenumbers and its logarithmic dependence on system size. In three dimensions, $\eta_R$ is finite in the infinite-volume limit and exhibits a minimum $\eta_R^{\min}$ as a function of the inverse (thermal) Reynolds number $\text{Re}_{\Lambda}^{-1}=\eta/\sqrt{\rho T\Lambda}$. Both perturbative and RG improved calculations agree with the direct simulations for large inverse (thermal) Reynolds number. In the regime of small inverse (thermal) Reynolds number, the RG improved result stays accurate down to the minimum of the renormalized viscosity $\eta_R^{\min} \approx 0.25\sqrt{\rho T \Lambda}$. Self-consistent RG improved calculations also capture the subsequent rise at small bare viscosities to within about $15\%$, whereas one-loop perturbation theory fails in this non-perturbative regime.

Within an EFT picture, this minimum has a natural interpretation. For a given measured $\eta_R$, it implies an upper bound $\Lambda_{\max}$ on the cutoff, beyond which no choice of bare viscosity reproduces the measured value. Beyond this scale, the stochastic Navier--Stokes equation ceases to be a valid description, and one should resort to a more fine-grained description.

Beyond our present studies a natural first extension is heat conduction in an incompressible fluid, where an additional diffusive energy (or entropy) density is advected by the transverse momentum through the Poisson bracket. Replacing the Gaussian Hamiltonian of this conserved order parameter field by a Landau--Ginzburg--Wilson functional then leads directly to Model~H for the critical dynamics of the liquid-gas transition, where non-Gaussian statics make the inclusion of a static regulator $R_\Lambda$ essential. On the numerical side, multiplicative noise from the fluctuating entropy and the resulting stiffness from fast dissipative UV modes call for new integrators, for instance implicit schemes.

Naturally, other important applications of this formalism concern the dynamics of compressible fluids as well as relativistic fluctuating hydrodynamics in the density frame~\cite{Armas:2020mpr,Basar:2024qxd,Bhambure:2024axa,Bhambure:2024gnf}. Dropping the incompressibility constraint leads to the appearance of sound modes, and the non-Gaussian static fluctuations of compressible fluids plausibly make the static regulator $R_\Lambda$ indispensable here as well.

Eventually,  as sketched in Fig.~\ref{fig:scales} the ultimate goal should be to 
match the bare parameters $\eta(\Lambda)$ to an underlying microscopic theory, which would make the EFT interpretation quantitative.

\acknowledgments

We thank Victor E. Ambru\c{s}, Lubomir Banas, Thomas Sch\"afer, Leon Sieke, Lorenz von Smekal, Derek Teaney and Yunxin Ye for insightful discussions.
This work was supported by the Deutsche Forschungsgemeinschaft (DFG, German Research Foundation) through the CRC-TR 211 ‘Strong-interaction matter under extreme conditions’-project number 315477589 – TRR 211. The authors gratefully acknowledge the computing time made available to them on the high-performance computer Otus at the NHR Center Paderborn Center for Parallel Computing (PC2). This center is jointly supported by the Federal Ministry of Research, Technology and Space and the state governments participating in the National High-Performance Computing (NHR) joint funding program (\url{https://www.nhr-verein.de/en/our-partners}).
\appendix

\section{\boldmath Fokker--Planck equation and stationarity of \texorpdfstring{$p_\text{eq}[\vec{\pi}]$}{the equilibrium distribution}} \label{sec:fokker_planck}

Since the noise in the equations of motion~\eqref{eq:eom_general} is additive, Gaussian, and white in time, no Itô--Stratonovich ambiguity arises, and the probability density functional $p_t[\vec{\pi}]$ of observing a configuration $\vec{\pi}$ at time $t$ evolves according to the Fokker--Planck equation \cite{taeuber_2014}
\begin{align}
    \partial_t p_t[\vec{\pi}]
    = &-\int_{\vec{x}}\frac{\delta}{\delta\pi_i(\vec{x})}\Big\{\Big(\partial_t\pi_i(\vec{x})\big|_\text{diss.}+\partial_t\pi_i(\vec{x})\big|_\text{rev.}\Big)\,p_t[\vec{\pi}]\Big\} \nonumber\\
    &-T\int_{\vec{x}\vec{y}}\frac{\delta}{\delta\pi_i(\vec{x})}\bigg\{\Gamma_{ij}(\vec{x}-\vec{y})\,\frac{\delta p_t[\vec{\pi}]}{\delta\pi_j(\vec{y})}\bigg\}, \label{eq:FP}
\end{align}
where
\begin{align}
    \partial_t\pi_i(\vec{x})\big|_\text{diss.} = \int_{\vec{y}}\Gamma_{ij}(\vec{x}-\vec{y})\,\frac{\delta H[\vec{\pi}]}{\delta\pi_j(\vec{y})}
\end{align}
denotes the dissipative drift and $\partial_t\pi_i(\vec{x})|_\text{rev.}$ the reversible drift given in Eq.~\eqref{eq:rev_eom}. The last term describes diffusion in configuration space with the positive semi-definite diffusion kernel $-T\Gamma_{ij}$, which is fixed by the FDT~\eqref{eq:FDT_general} in terms of the same dissipative kernel.\footnote{Since $\vec{\pi}$ is transverse, the configuration space is the space of transverse fields, and all functional derivatives are taken on this subspace, $\delta\pi_m(\vec{y})/\delta\pi_l(\vec{x}) = \mathcal{T}_{ml}(\vec{y}-\vec{x})$. In particular, the outer functional derivatives in Eq.~\eqref{eq:FP} only probe the transverse part of the terms they act on, such that no explicit projectors are needed in the dissipative and diffusive terms.}

We now show that $p_\text{eq}[\vec{\pi}]\propto e^{-H[\vec{\pi}]/T}$ is stationary, treating the dissipative-fluctuating and reversible contributions separately and ignoring the momentum constraint in Eq.~\eqref{eq:p_eq} for now. Using ${\frac{\delta p_\text{eq}[\vec{\pi}]}{\delta\pi_j(\vec{y})} = -\frac{1}{T}\frac{\delta H[\vec{\pi}]}{\delta\pi_j(\vec{y})}\,p_\text{eq}[\vec{\pi}]}$, the diffusion term cancels against the dissipative drift term already at the level of the probability current, i.e., the probability current of the dissipative-fluctuating part vanishes identically (detailed balance). This cancellation only relies on the FDT and holds for any choice of $H$ and $\Gamma$, provided that the same $H$ enters both the thermodynamic force and $p_\text{eq}$. Hence, $p_\text{eq}$ is stationary under the dissipative-fluctuating part of the dynamics. The remaining reversible contribution yields
\begin{align}
    \partial_t p_\text{eq}
    &= -\int_{\vec{x}}\frac{\delta}{\delta\pi_i(\vec{x})}\Big(p_\text{eq}\,\partial_t\pi_i(\vec{x})\big|_\text{rev.}\Big) \nonumber\\
    &= -p_\text{eq}\left[\int_{\vec{x}}\frac{\delta\,\partial_t\pi_i(\vec{x})|_\text{rev.}}{\delta\pi_i(\vec{x})} - \frac{1}{T}\,\partial_t H\big|_\text{rev.}\right], \label{eq:FP_rev}
\end{align}
where the first line is, up to a sign, the left-hand side of Eq.~\eqref{eq:divFreeCond}. The second term in square brackets vanishes for any $H$ due to conservation of $H$ under the reversible dynamics, as shown in Sec.~\ref{sec:rev_terms}. Hence, $p_\text{eq}$ is stationary under the reversible dynamics if and only if the reversible drift is divergence-free in configuration space, i.e., conserves phase-space volume, which is the condition~\eqref{eq:divFreeCond}.

For the regularized Hamiltonian~\eqref{eq:h_reg}, a lengthy but straightforward calculation, inserting Eqs.~\eqref{eq:rev_eom} and \eqref{eq:fluid_vel_reg} and evaluating the resulting kernels in Fourier space, shows that this is indeed the case, independently of $R_\Lambda$. The only non-trivial step concerns a contribution proportional to the total momentum $\vec{\Pi}$, which vanishes by parity in $\vec{p}$ and thus requires a parity-symmetric evaluation of a formally divergent wavenumber integral not regulated by $R_\Lambda$ or $K_\Lambda$. Hence, Eq.~\eqref{eq:divFreeCond} is satisfied.

Finally, including the momentum constraint ${\delta(\vec{\Pi}-\int_{\vec{x}}\vec{\pi})}$ in Eq.~\eqref{eq:p_eq} does not change this conclusion. Functional derivatives of the constraint only enter contracted with the zero-mode components of the drift and of the diffusion kernel. Both vanish, since $\Gamma_{ij}(\vec{p}=0)=0$ and the reversible dynamics conserves the total momentum (cf.~Eq.~\eqref{eq:Pi_conserv}). Altogether, $\partial_t p_\text{eq}=0$. Moreover, since the noise excites every mode except the conserved zero mode, $p_\text{eq}$ is the only stationary distribution at fixed $\vec{\Pi}$, such that the system relaxes to it from arbitrary initial conditions.

On the lattice, the dissipative-fluctuating step integrates the linear evolution of each mode exactly. Hence, a thermal ensemble remains thermal after each step, i.e., the step leaves $p_\text{eq}$ invariant for any $\Delta t$. The spatially discretized reversible dynamics~\eqref{eq:rev_lattice} conserves $H_\Lambda$ due to the antisymmetry of the discrete bracket~\eqref{eq:disc_bracket}, and its divergence vanishes by the same arguments as in the continuum, with $p_m$ replaced by $p^c_m$ and integrals by finite sums. In particular, the parity argument applies exactly on the lattice, since $p^c_m$ is odd in $\vec{p}$. Together with exact momentum conservation, the reversible step therefore also leaves $p_\text{eq}$ invariant, up to the error of the time integration.

\begin{widetext}

\section{Regulated one-loop correction to viscosity}
\label{sec:one_loop_calc}

We convert the (regulated) Langevin equation~\eqref{eq:reg_eom} into a path integral using the Martin-Siggia-Rose-Janssen-De Dominicis (MSRJD or often simply MSR) formalism \cite{Martin:1973zz,Dominicis:1976,Janssen:1976}. In this approach, correlation functions of $\vec{\pi}$ and $\tilde{\vec{\pi}}$ are obtained from the generating functional
\begin{align}
    Z[\vec{A},\tilde{\vec{A}}]& = \int \mathcal{D}\pi\,\mathcal{D}\tilde{\pi}\,\exp\bigg\{iS[\vec{\pi},\tilde{\vec{\pi}}] + i\int_{\Dvec{x}} \left( \vec{A}(\Dvec{x})\cdot\tilde{\vec{\pi}}(\Dvec{x}) +  \tilde{\vec{A}}(\Dvec{x})\cdot \vec{\pi}(\Dvec{x}) \right)\bigg\}
\end{align}
with
\begin{align}
    &S[\vec{\pi},\tilde{\vec{\pi}}] = \nonumber \\
    &\int_{\omega\vec{p}}
    \frac{1}{2}
    \begin{pmatrix}
        \pi_l(-\omega,-\vec{p}) &
        \tilde{\pi}_l(-\omega,-\vec{p})
    \end{pmatrix}
    \begin{pmatrix}
        0& -i\omega + \Gamma_{\Lambda}(p) \mathcal{T}_{lm}(\vec{p})(1+R_{\kk\Lambda}(p))/\rho \\
        i\omega + \Gamma_{\Lambda}(p) \mathcal{T}_{lm}(\vec{p})(1+R_{\kk\Lambda}(p))/\rho & -2iT \Gamma_{\Lambda}(p) \mathcal{T}_{lm}(\vec{p})
    \end{pmatrix}
    \begin{pmatrix}
        \pi_m(\omega,\vec{p}) \\
        \tilde{\pi}_m(\omega,\vec{p})
    \end{pmatrix} \nonumber \\
    &-\frac{g}{\rho}  \int_{t\vec{x}\vec{y}} \,
    \tilde{\pi}_l(t,\vec{x})\mathcal{T}_{lm}(\vec{x}-\vec{y}) \left(\pi_n(t,\vec{y}) +  \int_{\vec{z}} R_{\kk\Lambda}(\vec{y}-\vec{z}) \pi_n(t,\vec{z})\right)
      \left(\frac{\partial \pi_m(t,\vec{y})}{\partial y_n}- \frac{\partial \pi_n(t,\vec{y})}{\partial y_m}\right)  \label{eq:eff_avg_action}
\end{align}
Here, $\Gamma_\Lambda(p) = -\eta p^2(1+K_\Lambda(p))$ denotes the regularized dissipative kernel, $R_{\kk\Lambda}=R_\kk+R_\Lambda$, and $g$ is the mode coupling constant, which is fixed to $g=1$ by Galilean invariance but kept explicit to count the order of perturbation theory.

\begin{figure*}[t]
\centering
\begin{tikzpicture}[
    pif/.style={smooth},
    pitf/.style={smooth,densely dotted},
    mom/.style={blue,-latex},
    lab/.style={blue,font=\footnotesize},
    lhs/.style={anchor=base east,inner xsep=0pt},
    rhs/.style={anchor=base west,inner xsep=0pt,xshift=0.5em},
  ]

  \begin{scope}[shift={(0,0)}]
    \draw[pif] plot[domain=0:2,samples=400,variable=\x] (\x,{0.12*sin(1440*\x)});
    \node[lab,left]  at (0,0) {$l$};
    \node[lab,right] at (2,0) {$m$};
    \draw[mom] (0.6,0.32) -- (1.4,0.32);
    \node[lab] at (1,0.58) {$\omega,\vec{p}$};
    \node[lhs] at (5.4,-0.08) {$G_{\kk(0)}^{\pi_l \pi_m}(\omega,\vec{p}) = {}$};
    \node[rhs] at (5.4,-0.08) {$\displaystyle
      -\frac{2iT \Gamma_{\Lambda}(p)\,\mathcal{T}_{lm}(\vec{p})}
           {\omega^2 + \big(\Gamma_{\Lambda}(p)(1+R_{\kk\Lambda}(p))/\rho\big)^2}$};
  \end{scope}

  \begin{scope}[shift={(0,-1.6)}]
    \draw[pif]  plot[domain=0:1,samples=200,variable=\x] (\x,{0.12*sin(1440*\x)});
    \draw[pitf] plot[domain=1:2,samples=200,variable=\x] (\x,{0.12*sin(1440*\x)});
    \node[lab,left]  at (0,0) {$l$};
    \node[lab,right] at (2,0) {$m$};
    \draw[mom] (0.6,0.32) -- (1.4,0.32);
    \node[lab] at (1,0.58) {$\omega,\vec{p}$};
    \node[lhs] at (5.4,-0.08) {$G_{\kk(0)}^{\pi_l \tilde{\pi}_m}(\omega,\vec{p}) = {}$};
    \node[rhs] at (5.4,-0.08) {$\displaystyle
      -\frac{\mathcal{T}_{lm}(\vec{p})}{i\omega + \Gamma_{\Lambda}(p)(1+R_{\kk\Lambda}(p))/\rho}$};
  \end{scope}

  \begin{scope}[shift={(0,-3.2)}]
    \draw[pitf] plot[domain=0:1,samples=200,variable=\x] (\x,{0.12*sin(1440*\x)});
    \draw[pif]  plot[domain=1:2,samples=200,variable=\x] (\x,{0.12*sin(1440*\x)});
    \node[lab,left]  at (0,0) {$l$};
    \node[lab,right] at (2,0) {$m$};
    \draw[mom] (0.6,0.32) -- (1.4,0.32);
    \node[lab] at (1,0.58) {$\omega,\vec{p}$};
    \node[lhs] at (5.4,-0.08) {$G_{\kk(0)}^{\tilde{\pi}_l \pi_m}(\omega,\vec{p}) = {}$};
    \node[rhs] at (5.4,-0.08) {$\displaystyle
      -\frac{\mathcal{T}_{lm}(\vec{p})}{-i\omega + \Gamma_{\Lambda}(p)(1+R_{\kk\Lambda}(p))/\rho}$};
  \end{scope}

  \begin{scope}[shift={(0,-4.5)}]
    \draw[pitf] plot[domain=0:2,samples=400,variable=\x] (\x,{0.12*sin(1440*\x)});
    \node[lab,left]  at (0,0) {$l$};
    \node[lab,right] at (2,0) {$m$};
    \draw[mom] (0.6,0.32) -- (1.4,0.32);
    \node[lab] at (1,0.58) {$\omega,\vec{p}$};
    \node[lhs] at (5.4,-0.08) {$G_{\kk(0)}^{\tilde{\pi}_l \tilde{\pi}_m}(\omega,\vec{p}) = {}$};
    \node[rhs] at (5.4,-0.08) {$0$};
  \end{scope}

  \begin{scope}[shift={(1.1,-6.35)}]
    \draw[pitf] plot[domain=-1:0,samples=200,variable=\x] (\x,{0.12*sin(1440*\x)});
    \node[lab,left] at (-1,0) {$l$};
    \draw[mom] (-0.9,0.32) -- (-0.3,0.32);
    \node[lab] at (-0.6,0.58) {$\Dvec{p}$};
    \begin{scope}[rotate=40]
      \draw[pif] plot[domain=0:1,samples=200,variable=\x] (\x,{0.12*sin(1440*\x)});
      \node[lab,above right] at (1,0) {$m$};
      \draw[mom] (0.9,0.32) -- (0.3,0.32);
      \node[lab] at (0.6,0.58) {$\Dvec{q}$};
    \end{scope}
    \begin{scope}[rotate=-40]
      \draw[pif] plot[domain=0:1,samples=200,variable=\x] (\x,{0.12*sin(1440*\x)});
      \node[lab,below right] at (1,0) {$n$};
      \draw[mom] (0.9,-0.32) -- (0.3,-0.32);
      \node[lab] at (0.6,-0.58) {$\Dvec{r}$};
    \end{scope}
    \fill (0,0) circle (0.05);
  \end{scope}
  \node[lhs] at (5.4,-6.05) {$S_\kk^{\tilde{\pi}_l \pi_m \pi_n}(\Dvec{p},\Dvec{q},\Dvec{r}) = {}$};
  \node[rhs] at (5.4,-6.05) {$\begin{aligned}[t]
    &{-}\frac{g}{\rho} \Big\{
      \mathcal{T}_{lm}(\vec{p})\, i q_n (1+R_{\kk\Lambda}(r))
      + \mathcal{T}_{ln}(\vec{p})\, i r_m (1+R_{\kk\Lambda}(q)) \\
    &\qquad - \mathcal{T}_{lk}(\vec{p})
      \big[ i q_k (1+R_{\kk\Lambda}(r)) + i r_k (1+R_{\kk\Lambda}(q)) \big] \delta_{mn}
      \Big\}\, (2\pi)^{d+1} \delta(\Dvec{p}+\Dvec{q}+\Dvec{r})
  \end{aligned}$};

\end{tikzpicture}
\caption{Feynman rules for the MSR action \eqref{eq:eff_avg_action}. \label{fig:feynmanRules}}
\end{figure*}

The one-loop correction to the (retarded) self energy $\Sigma^{\tilde{\pi}\pi}$ is most conveniently represented diagrammatically (using the rules from Fig.~\ref{fig:feynmanRules}),
\begin{align}
    \Sigma_\kk^{\tilde{\pi}_l \pi_m}(\omega=0,\vec{p}) &=
    i\;\; \begin{tikzpicture}[baseline=-0.5ex]
        \draw[] plot[domain=0:180, samples=200, smooth, variable=\a] ({0.5*cos(\a)*(1.0-0.1*sin(12*\a))},{0.5*sin(\a)*(1.0-0.1*sin(12*\a))});
        \draw[blue,->] (70:0.7) arc (70:110:0.7);
        \node[blue] at (0,+0.95) {\footnotesize $(\nu,\vec{q})$};
        \draw[] plot[domain=180:270, samples=200, smooth, variable=\a] ({0.5*cos(\a)*(1.0-0.1*sin(12*\a))},{0.5*sin(\a)*(1.0-0.1*sin(12*\a))});
        \draw[densely dotted] plot[domain=270:360, samples=200, smooth, variable=\a] ({0.5*cos(\a)*(1.0-0.1*sin(12*\a))},{0.5*sin(\a)*(1.0-0.1*sin(12*\a))});
        \draw[blue,->] (-70:0.7) arc (-70:-110:0.7);
        \node[blue] at (0,-0.95) {\footnotesize $(-\nu,\vec{p}-\vec{q})$};
		\draw[densely dotted,smooth,domain=0:0.5,variable=\x] plot (-\x-0.5,{-0.05*sin(8*3.14159*\x r)});
        \draw[blue,->] (-0.7,0.2) -- (-1.1,0.2);
        \node[blue] at (-1.0,+0.45) {\footnotesize $(0,\vec{p})$};
        \node[blue] at (-0.9,-0.25) {\footnotesize $l$};
        \draw[smooth,domain=0:0.5,variable=\x] plot (\x+0.5,{0.05*sin(8*3.14159*\x r)});
        \draw[blue,->] (1.1,0.2) -- (0.7,0.2);
        \node[blue] at (1.0,+0.45) {\footnotesize $(0,\vec{p})$};
        \node[blue] at (+0.9,-0.3) {\footnotesize $m$};
	 	\fill[black] (-0.5,0) circle (0.05);
	 	\fill[black] (+0.5,0) circle (0.05);
    \end{tikzpicture}
\end{align}

We evaluate the index contractions using Mathematica \cite{Mathematica} and solve the frequency ($\nu$) integral with the residue theorem, closing the contour in lower half plane and picking up the first-order poles of $G_{\kk(0)}^{\pi\pi}(\nu,\vec{q})$ and $G_{\kk(0)}^{\pi\pi}(-\nu,\vec{p}-\vec{q})$ at $\nu = i\Gamma(q)(1+R_{\kk}(q)) /\rho$ and $\nu = i\Gamma(|\vec{p}-\vec{q}|)(1+R_{\kk}(|\vec{p}-\vec{q}|))/\rho$, respectively. Subsequently, we write the remaining integral over $\vec{q}$ as an integral over $q=|\vec{q}|$ and $u=\cos\theta$ in hyperspherical coordinates (where $\theta$ denotes the angle between $\vec{q}$ and $\vec{p}$). We obtain the result
\begin{align}
    \Sigma_\kk^{\tilde{\pi}\pi}(\omega=0,p) &= \frac{S_{d-2} g^2 \rho  T}{(2\pi)^d (d-1)} \int_{0}^{\infty} dq \,q^{d-1} \int_{-1}^{1}du \,(1-u^2)^{(d-1)/2}   \times \nonumber\\
    &\frac{ 1 }{r^2 \left(R_{\kk\Lambda }\left(q\right)+1\right) \left(-\Gamma_{\Lambda} \left(q\right) \left(R_{\kk\Lambda }\left(q\right)+1\right)-\Gamma_{\Lambda} \left(r\right) \left(R_{\kk\Lambda }\left(r\right)+1\right)\right)} \times \nonumber \\
    &{\color{red}\Bigg\{}(d-1) p^4 \left(R_{\kk\Lambda }\left(q\right)+1\right){}^2+p^2 q {\color{green}\bigg[}q {\color{blue}\Big(} R_{\kk\Lambda }\left(p\right) \left(-(d-1) R_{\kk\Lambda }\left(q\right)+2 (d-2) R_{\kk\Lambda }\left(r\right)+d-3\right) \nonumber \\
    &-(d-1) R_{\kk\Lambda }\left(q\right) R_{\kk\Lambda }\left(r\right)+(d-1) R_{\kk\Lambda }\left(q\right){}^2+(d-3) \left(R_{\kk\Lambda }\left(r\right)+1\right){\color{blue}\Big)}-2 (d-1) p u \left(R_{\kk\Lambda }\left(q\right)+1\right){}^2{\color{green}\bigg]} \nonumber \\
    &+q^3 \left(R_{\kk\Lambda }\left(p\right)+1\right) {\color{green}\bigg[}-(d-1) (q-2 p u) R_{\kk\Lambda }\left(q\right)+R_{\kk\Lambda }\left(r\right) {\color{blue}\Big(}(d-1) q-2 (d-2) p u{\color{blue}\Big)}+2 p u{\color{green}\bigg]}{\color{red}\Bigg\}} \label{eq:p_dep_self_en}
\end{align}
The correction to the viscosity can be extracted from the $p\to 0$ behavior of the self energy (cf.~Eq.~\eqref{eq:eta_corr_from_self_en} in the main text),\footnote{In practice, differentiating the integrand in Eq.~\eqref{eq:p_dep_self_en} w.r.t. $p^2$ and expanding around $p=0$ yields divergent terms proportional to $1/p$ which are odd in $u$ and thus vanish upon the $u$ integration.}
\begin{align}
    \Delta \eta_\kk = -\lim_{p^2 \to 0} \frac{\partial}{\partial p^2}\frac{\rho\Sigma^{\tilde{\pi}\pi}_\kk(\omega=0,p)}{(1+R_{\kk\Lambda}(p))(1+K_{\Lambda}(p))}
\end{align}
After taking the limit $p^2\to 0$ the $u$ integral can be solved analytically. Inserting the bare dissipative kernel $\Gamma_\Lambda(p) = -\eta p^2(1+K_{\Lambda}(p))$ on the right-hand side and setting the IR regulator $R_{\kk}(p)=0$ to zero leads to Eq.~\eqref{eq:Del_eta} from the main text.

\section{Numerical implementation of RG improved flow equation}
\label{sec:rFRG_num_impl}

We perform the integral over $u\in[-1,1]$ using a Gauss-Gegenbauer quadrature adapted to the integration measure $(1-u^2)^{(d-1)/2}du$ in Eq.~\eqref{eq:p_dep_self_en}, and the integral over $q\in[0,\infty)$ using a Gauss-Legendre quadrature in $\ln(q/\kk)\in(-\infty,\infty)$. Both quadratures are provided by \cite{GSL}. In all our results we have explicitly verified that both the interval size (for $\ln(q/\kk)$) and the number of nodes are sufficiently large. We integrate the resulting set of ordinary differential equations in RG time $\ln(\kappa/\Lambda)$ using an adaptive Runge--Kutta--Fehlberg scheme provided by \cite{GSL}. In practice, typical values of the UV and IR cutoffs are $\kk_\text{UV}/\Lambda \sim 10^5$ and $\kk_\text{IR}/\Lambda \sim 10^{-3}$. %

\end{widetext}

\end{document}